%% file: AGN-SMBH-paper.tex
\documentclass[preprint2]{aastex}

\usepackage[squaren]{SIunits}

\usepackage{ wasysym }
\usepackage{graphicx}
\usepackage{url}
\usepackage{epstopdf}

\usepackage{pdflscape}

\newcommand {\HeII}  {\ion{He}{2}}    

\newcommand {\MgII}    {\ion{Mg}{2}}  

\newcommand {\OVI}    {\ion{O}{6}}   

\newcommand {\CII}     {\ion{C}{2}}   
\newcommand {\CIV}    {\ion{C}{4}}  

\newcommand {\HST}    {{\it HST}}

\slugcomment{Accepted to the Astrophysical Journal on  19 July 2013.}

\shorttitle{AGN Emission-SMBH Correlations}
\shortauthors{Tilton \& Shull}

\begin{document}
\defcitealias{nistasd}{NIST Atomic Spectra Database}


\title{Ultraviolet Emission-Line Correlations in Hubble/COS Spectra of \\ Active Galactic Nuclei:  Single-Epoch Black Hole Masses\footnote{Based on observations made with the NASA/ESA \textit{Hubble Space Telescope}, obtained from the data archive at the Space Telescope Science Institute. STScI is operated by the Association of Universities for Research in Astronomy, Inc., under NASA contract NAS5-26555.}}

\author{Evan M. Tilton and J. Michael Shull}
\affil{CASA, Department of Astrophysical and Planetary Sciences, University of Colorado, 389-UCB, Boulder, CO 80309}
\email{evan.tilton@colorado.edu, michael.shull@colorado.edu}

\begin{abstract}
Effective methods of measuring supermassive black hole masses in active galactic nuclei (AGN) are of critical importance to studies of galaxy evolution. While there has been much success obtaining masses through reverberation mapping, the extensive observing time required by this method has  limited the practicality of applying it to large samples at a variety of redshifts. This limitation highlights the need to estimate these masses using single-epoch spectroscopy of ultraviolet emission lines.  We use ultraviolet spectra of 44 AGN from \textit{HST}/COS, \textit{IUE}, and \textit{FUSE}  of the \ion{C}{4} $\lambda 1549$, \ion{O}{6} $\lambda 1035$, \ion{O}{3}] $\lambda 1664$, \HeII\ $\lambda1640$, \CII\ $\lambda 1335$, and \ion{Mg}{2} $\lambda 2800$ emission lines and explore their potential as tracers of the broad-line region and supermassive black hole mass.  The higher S/N and better spectral resolution of the Cosmic Origins Spectrograph on {\it Hubble Space Telescope} resolves AGN intrinsic absorption and produces more accurate line widths.  From these, we test the viability of  mass-scaling relationships based on line widths and luminosities and carry out a principal component analysis based on line luminosities, widths, skewness, and kurtosis.  At $L_{1450} \leq 10^{45}$ erg~s$^{-1}$, the UV line luminosities correlate well with H$\beta$, as does the 1450~\AA\ continuum luminosity.   We find that \CIV, \OVI, and \MgII\ can be used as reasonably accurate estimators of AGN black hole masses, while \HeII\ and \CII\ are uncorrelated.    

\end{abstract}


\defcitealias{ves06}{VP06}

\bibpunct[ ]{(}{)}{;}{a}{}{,}

\clearpage
\section{INTRODUCTION}

The prevailing view of the structure of massive galaxies with hot spheroidal stellar components (bulges) places supermassive black holes (SMBHs) with masses in excess of $10^6~M_{\astrosun}$ at their centers \citep[for a review, see][]{fer05}.  The well-observed correlations of SMBH mass with host galaxy properties such as stellar velocity dispersion and $M-\sigma$ relation \citep{fer00,geb00,gul09,mcc11}, bulge mass \\ \citep[]{kor95}, and luminosity \citep{marc03} imply a co-evolution of SMBHs and galaxies, making SMBH growth a major research topic in modern cosmology. Recent numerical models within a hierarchical $\Lambda$CDM framework have begun to reproduce the observed SMBH population \citep[e.g.,][]{sha09,she09}, but unresolved questions remain, such as the source of black hole seeds, their growth rates, and feedback mechanisms.

These SMBHs are thought to build up during a phase of rapid gas accretion \citep[e.g.,][]{lyn69} onto objects we observe as quasars or active galactic nuclei (AGN). The accretion disks that form around the black holes release large amounts of radiative energy, which in turn photoionizes surrounding gas, observed as emission-line spectra superimposed on top of the disks' continuum spectra and referred to as the broad line region (BLR). These emission lines can display velocity widths upwards of \unit{10,000}{\kilo\meter~\reciprocal\second}, although they are more typically \unit{1000-5000}{\kilo\meter~\reciprocal\second}. Numerous authors have characterized the luminosity function (LF) of the AGN population at a variety of redshifts and in different wavelength regimes \citep[e.g.,][]{sch83,has05,ric05,ric06,hop07,cro09,air10,gli11}. Based on such surveys, the space density of quasars appears to have been much higher in the past, peaking $z \approx 2-3$.  

Using a line of reasoning from \citet{sol82}, one can estimate the total SMBH-accreted mass based on the total observed luminosity.  The AGN population in the past seems to roughly agree with modern, dormant SMBH population. This general agreement further solidifies the connection between these populations. A complete understanding of galaxy evolution therefore requires a detailed understanding of the demographics of the AGN population as a function of redshift, and it would be greatly improved if astronomers developed reliable estimates of SMBH masses out to $z = 2-3$.  While the LF is an important constraint on models of SMBH growth and evolution, the mass function for SMBHs more directly traces growth and accretion efficiency. Unfortunately, the mass function is not as well known as the LF, owing partially to the inherent difficulty in measuring SMBH masses \citep[for recent attempts to measure the mass function, see][]{kelly2010,kelly2013}.

Direct dynamical measurements are not yet possible for most active SMBHs, with such techniques being limited to only a handful of the most nearby quiescent SMBHs where velocity measurements can be performed near the black hole's sphere of influence ($R_{\rm BH}\sim GM_{\rm BH}/\sigma^2$) where $\sigma$ is the velocity dispersion of the central cusp \citep{bin08}.  Other, more indirect methods for mass estimation have been developed. These methods focus on the behavior of the emission properties of the BLR, assumed to trace the SMBH potential. Velocity-resolved reverberation mapping \citep{bla82,gas88,pet93,gas09} allows the measurement of the radius of the BLR ($R_{\rm BLR}$) through the observation of the time lag ($\tau$) of the response of a BLR emission line to variations in continuum luminosity, $L$, from the accretion disk through time-series spectral observations. Related results suggest that motions in the BLR are gravitationally dominated, and that these motions are traced by both high- and low-ionization lines \citep{pet99,pet00}.  Thus, reverberation mapping offers a method of SMBH mass determination in AGN.

 Mass estimation through reverberation mapping can be applied at any redshift or luminosity. Unfortunately, it is often impractical, owing to its observationally intensive requirements, which can be exacerbated by the long variability timescales of time-dilated, high-luminosity objects at high redshift. Thus, alternate methods of mass determination are required to build up a large sample of active SMBH masses \citep[see, e.g.,][for a review]{mar12}. Empirical results have shown that the radius of the BLR scales with the nuclear continuum luminosity as $R_{\rm BLR}\propto L^b$, implying that mass estimates can be made with single-epoch spectra, using continuum luminosity to trace $R_{\rm BLR}$ and width of the BLR component of emission lines ($\Delta v$) to trace the gravitational motion of the emitting clouds. One obtains a a virial estimate \citep[e.g.,][]{ves06} of the mass of the SMBH ($M_{\rm BH}$) by assuming
\begin{eqnarray}
M_{\rm BH}=f\frac{R_{\rm BLR}(\Delta v)^2}{G}  \;  , 
\end{eqnarray}
where $R_{\rm BLR}=c\tau\propto L^\beta$, $G$ is the gravitational constant, and $f$ is a scale factor of order unity that depends on the (unknown) geometry, structure, and inclination of the BLR \citep{onk04,pet04}. One expects $\beta\sim 0.5$ if gas densities and ionization parameters are similar across AGN and continuum shapes do not substantially vary with luminosity \citep[e.g.,][and references therein]{wan85}. Empirical values for $\beta$ mostly agree with these predictions, but there may be some variation depending on the luminosity diagnostic used \citep{kas00,kas05,ben06,ben09,ben13}.

These single-epoch virial relationships have been empirically calibrated by fitting the parameters to give masses that match those from the reverberation mapped sample, which in turn is calibrated to the $M-\sigma$ relation. In the optical regime, the full-width-at-half-maximum (FWHM) of H$\beta$ used with the monochromatic luminosity at 5100 \AA ~($L_{5100}$) is the most-studied single-epoch relationship. These H$\beta$ scaling relationships yield masses that are consistent on average with those obtained through H$\beta$ reverberation mapping, but with a scatter of $\sim 0.4$ dex \citep{mcl02,ves06}.  Because H$\beta$ shifts out of the optical for $z\gtrsim0.8$ and is therefore inaccessible to many ground-based observatories, it becomes a more difficult diagnostic to use for surveys at high redshift. Consequently, authors have calibrated several ultraviolet (UV) lines, most notably the doublets of  \ion{C}{4} $\lambda\lambda 1548,1551$ and \ion{Mg}{2} $\lambda\lambda 2796,2803$, along with alternate luminosities such as $L_{1350}$, $L_{1450}$, $L_{3000}$, and the luminosities of the lines themselves \citep{mcl02,ves02,wan04,ves06,onk08,ves09}. These diagnostics are even further removed from a direct mass measure because they are typically calibrated against the H$\beta$ reverberation mapping estimates, owing to a paucity of reverberation measurements using UV lines. However, more results derived from observations of \ion{Mg}{2} and \ion{C}{4} are slowly becoming available \citep[see][for a review]{pet11}. While the \ion{Mg}{2}-based estimates correlate well with the Balmer line estimates \citep{sal07,mcg08,onk08,she08,wan09,she12}, the \ion{C}{4}-based estimates have proven far more contentious. 

Because the \ion{Mg}{2} line moves out of the optical at $z>2$, it would be useful if high-ionization lines in the far UV (FUV) could be used to estimate masses reliably. However, the issue is complicated by  the many ways in which the phenomenology of high-ionization lines differs from that of low-ionization lines \citep[for a review, see][]{sul00}. Foremost among the complications is the observed blueshift of the \ion{C}{4} line relative to the galaxy's systemic velocity as gauged by \ion{Mg}{2} or H$\beta$ \citep[e.g.,][]{gas82,bia12}. Numerous authors further claim that high-ionization lines are preferentially asymmetric and that their widths may not correlate with low-ionization lines, especially at higher luminosities \citep[e.g.,][]{wil93,wil95,bas05,net07,sul07,she08}. \citet{she12}, for example, analyzed a set of 60 quasars at $z\sim1.5-2.2$ from the SDSS DR7 quasar catalog. They compared the widths of the various UV lines, including \ion{C}{4}, with measurements of the Balmer lines in ground-based near infrared spectra, but found nearly no correlation between the two sets of lines. It is difficult to reconcile such discrepancies if both sets of lines originate in the same gravitationally-dominated, BLR gas (though AGN variability may also contribute to their results). These results suggest that the high-ionization lines have a significant non-gravitationally dominated component, perhaps arising as outflow from a disk wind \citep[e.g.,][]{lei04,ric11,wan11}. In contrast, other authors find that single-epoch mass estimates using \ion{C}{4} are completely consistent with those made with low-ionization lines. \citet{ves06}, for example, used 27 AGN with space-based UV data from \textit{IUE}, the \textit{Hopkins Ultraviolet Telescope}, and \textit{HST} to calibrate \ion{C}{4} scaling relationships consistently with H$\beta$. \citet{kel07} confirmed these results with a similar sample of space-based observations, and \citet{ass11} found that mass estimates made with single-epoch \ion{C}{4} measurements are consistent with estimates made with H$\beta$ in high-redshift targets using a sample of lensed AGN.

One possible explanation for these discrepant results is that they are the effects of varying data quality. \citet{den13} suggest that much of the disagreement arises from the use of low-S/N data when measuring \ion{C}{4} line widths \citep[see also][for further discussion of this issue]{den11,den12,ves11,woo11}. Other recent work has attempted to reconcile these discrepant results by searching for a parameter to correct for non-gravitational components that may contribute to the line profile. Corrections based on the $L_{1350}/L_{5100}$ ratio \citep{ass11}, the relative \ion{C}{4} blueshift \citep{she08}, or the shape of the line profile \citep[especially with respect to kurtosis;][]{den12} have shown the most promise. However, it must be stressed that, regardless of the emission line used, virial-mass estimates are fundamentally different from true mass measurements. Virial relationships may be biased estimators that can lead to misleading results for the underlying mass distribution \citep{she08,kel09,she10,she12b}. Because of the large intrinsic scatter of the scaling relationship distribution, SMBH mass functions can be broadened, leading to an underestimated peak and overestimated tails. These biases are exacerbated in flux-limited samples, because a range of SMBH masses exists at a given luminosity, so a simple completeness correction in flux does not fully account for incompleteness in mass. Nonetheless, it is important to quantify the utility of single-epoch mass estimators, because they are currently the only viable way to estimate SMBH masses for large samples over a range of redshifts.

In this paper, we use UV spectra of 44 nearby bright AGN to directly measure the line profiles of several UV emission lines. We then address the viability of FUV emission lines as single-epoch virial estimators of active SMBH mass. The Cosmic Origins Spectrograph on the \textit{Hubble Space Telescope} (\textit{HST}) has now observed hundreds of AGN in the low-redshift universe with a greater sensitivity than any past UV mission. This provides a unique FUV dataset with both high signal-to-noise (S/N) and spectral resolution exceeding that of the spectra typically used for such studies by more than an order of magnitude. This allows us to remove narrow intervening absorption that may introduce errors in spectral measurements. Coupled with archival spectra from the \textit{Far Ultraviolet Spectroscopic Explorer} (\textit{FUSE}) and the \textit{International Ultraviolet Explorer} (\textit{IUE}), we are able to address the \ion{C}{4} discrepancy by testing the effects of data quality and additional line diagnostics on inferred single-epoch masses. We also investigate the potential of other FUV lines such as \ion{O}{6} $\lambda\lambda 1032,1038$ and \ion{O}{3}] $\lambda1664$ as tracers of SMBH mass. We test a wide variety of  line parameters, including alternate measures of line width and line shape, for correlations with black hole mass that may improve single-epoch mass estimators.

\section{METHODOLOGY}

In this section, we describe our sample and the measurements of numerous emission-line and continuum parameters that we investigate as tracers of SMBH mass. Throughout this paper we adopt atomic transition wavelengths and other atomic properties from \citet{mor03}. When that catalog does not contain the line of interest, we use the NIST Atomic Spectra Database.\footnote{\url{http://www.nist.gov/pml/data/asd.cfm}} We adopt a flat $\Lambda$CDM cosmology with $H_0=71~{\rm km~s^{-1}~Mpc^{-1}}$, $\Omega_m=0.27$, and $\Omega_\Lambda=0.73$.

\subsection{Object Selection and Physical Properties}
\label{sec:mass}


\input{agntable-2}

We selected targets with available COS data and mass estimates from reverberation mapping studies and/or single-epoch H$\beta$ measurements. These 44 objects and their properties are listed in Columns 1--4 of Table~\ref{tab:agn}.   Their redshifts are taken from the NASA/IPAC Extragalactic Database\footnote{\url{http://ned.ipac.caltech.edu}} (Column 2) and from our UV spectra (Column 3), using the peaks of our fits to the \ion{O}{3}] $\lambda\lambda 1660, 1666$ or Ly$\alpha$ $\lambda 1216$ emission lines, depending on line availability.  Random statistical error in the these redshift determinations is negligible, but the two redshift measurements often differ systematically, owing to the well-known offset  of high-ionization emission lines from systemic redshifts.  Because the sample was selected solely on the basis of the availability of data and measurements, it is comprised primarily of bright Seyfert galaxies, whose ionization subclasses are listed in Column 4.   The sample of 44 AGN includes four flat-spectrum radio quasars (3C\,273, PKS\,1302-102, Ton\,580, and Mrk\,1044), consistent with the typical 10\% radio-loud fraction.  Columns 5--12 give line widths, luminosities, and inferred SMBH masses.

Line widths and luminosities for the broad component of H$\beta$ were obtained from a variety of studies, primarily \citet{mar03}. When multiple studies reported on an object, we favored this study for the sake of homogeneity, as it covers the largest number of objects in our sample. It has the additional benefit of having been used in past work such as \citet{ves06}, allowing straightforward comparison. We corrected their line widths for resolution-broadening by using their reported spectral resolutions with the method described by \cite{pet04}, and we corrected their continuum fluxes for extinction assuming a \citet{fit99} reddening law with extinction values derived from \citet{sch11}. The continuum fluxes and equivalent widths (EW) were used to obtain rest-frame line luminosities. Although \citet{mar03} do not report errors on these quantities, we adopted 10\% error bars on the FWHM and EW measurements, which is roughly what they estimate the $2\sigma$ error would be on these measurements. We estimated the continuum error based on the quoted signal-to-noise ratio (S/N), though this may slightly underestimate the uncertainty arising from the flux calibration. We added an additional 10\% error to the EW and continuum measurements for objects flagged as uncertain in that study.  While the \citet{mar03} luminosities result from data of varying quality, often taken during non-photometric conditions, they appear to be consistent within intrinsic AGN variability \citep{ves06}. We followed similar procedures with other measurements. Line widths from \citet{goo89} and continuum luminosities from \citet{oht07} were used unchanged, because they were already corrected for instrumental resolution and extinction, respectively. Measurements from \citet{sti90} were corrected for resolution and extinction as described above. Measurements from \citet{gru99,gru04} were used unchanged. In all cases in which the authors did not report errors, we adopted 10\% error on the quantity. 

Throughout this paper, we adopt the single-epoch H$\beta$ mass estimates calculated using Equations~5 or 6 of \citet{ves06} as the reference black hole mass unless otherwise stated. The H$\beta$ emission-line width is well-established as a reliable tracer of black hole mass, with an intrinsic scatter of $\sim 0.4$ dex, although it may be less reliable for narrow-line Seyfert galaxies \citep{ves06,ass11}. We use the H$\beta$ calibrations from \citet{ves06} instead of the more recent calibration by \citet{ass11}, which uses the newer $R_{\rm BLR}-\lambda L_{5100}$ relation from \citet{ben09}, because the former allows self-consistent estimates with either $L(\rm H\beta )$ or $\lambda L_{5100}$. These calibrations yield mass estimates that are approximately 0.015 dex smaller than the newer calibration; the difference is negligible compared to measurement error and intrinsic scatter.   The black-hole mass estimates from H$\beta$ are reported in Column 9 of Table~\ref{tab:agn}, and Column 10  lists masses from reverberation mapping where available. For a few targets lacking single-epoch estimates, we adopt the reverberation results for the black hole mass. Single-epoch H$\beta$ mass estimates are expected to be systematically higher than reverberation results, owing to host galaxy starlight contamination, which varies with luminosity and slit size. As in other studies investigating the systematics of single-epoch mass estimates, we are unable to determine the size of this offset for objects in our sample. Host galaxy contamination is expected to be less significant for the UV luminosities discussed throughout this paper owing to the lower UV starlight luminosity of a typical galaxy. 
The types of data that were used for each object are listed in Column 12.

\subsection{UV Data}

For each object, we analyzed any available data taken with COS, \textit{FUSE}, and \textit{IUE}. Calibrated data from the COS/G130M ($1133-1468$~\AA) and COS/G160M ($1383-1796$~\AA) medium-resolution gratings ($R\equiv\lambda/\Delta\lambda\approx 18,000$) data were retrieved from the Mikulski Archive for Space Telescopes (MAST).\footnote{\url{http://archive.stsci.edu}} The properties of the COS instrument are described in detail by \citet{gre12} and \citet{ost11}. The separate exposures were coadded using the routines provided on the COS Tools website\footnote{\url{http://casa.colorado.edu/~danforth/science/cos/costools.html}} and described in detail by \citet{dan10} and \citet{kee12a}.

The \textit{FUSE} satellite consisted of four co-aligned telescopes leading to two detectors with two segments each, covering a total wavelength range of \unit{905-1187}{\angstrom} with $R\approx 20,000$ \citep{moo00}. In this work, we use only the LiF1A and LiF2A channels. This combination offers the highest throughput at the wavelengths of interest ($\lambda>\unit{1000}{\angstrom}$) while avoiding complications from the prominent, so-called ``worm'' feature, which arises from a grid-wire shadow above the detector and is most prominent in the 
LiF1B data. The calibrated  \textit{FUSE} exposures were retrieved from MAST before they were cross-correlated to account for small wavelength errors between the exposures and coadded.

The available IUE data vary greatly in quality and observing cadence. Some targets were observed with a few long exposures, and others were observed in reverberation mapping campaigns that resulted in several hundred separate exposures taken over several years. We retrieved all calibrated spectra taken in IUE's low-dispersion, large-aperture mode in the short-wavelength ($1150-2000$~\AA) and long-wavelength ($1850-3300$~\AA) channels; these spectra were extracted using the IUEDAC IDL routines.\footnote{\url{http://archive.stsci.edu/iue/iuedac.html}}  The resolving power varies considerably with wavelength, with typical values of $R=200 - 600$ \citep{hol82,cas83}.  Some targets featured several very low-S/N exposures that were several orders of magnitude discrepant in flux compared to the other exposures. We therefore visually inspected each exposure and discarded those with wildly discrepant flux levels compared to other exposures of the same object. The remaining exposures were coadded. 

Because the data for each object came from multiple instruments and were taken at different times, the continuum flux levels are not always consistent among the datasets, owing to source variation and/or inconsistent flux calibrations. To enable comparisons among the properties of emission-line and continuum features measured in the different spectra, we scale the \textit{FUSE} and \textit{IUE} spectra to the COS spectra by minimizing the differences in continuum regions of spectral overlap. These scale factors are all of order unity, with means of 0.99 and 0.91 for \textit{FUSE} and \textit{IUE} data, respectively. The discrepancy between the IUE and COS flux levels is small enough to be plausibly attributed to AGN variability \citep[see, e.g.,][]{wil05}. Differences in the flux calibrations of COS, thought to be accurate to better than $5\%$ \citep[initially reported in][]{mas10}, and IUE \citep{gar97} may also contribute to the discrepancy.  Owing to the various instruments, coadditions, and scalings used, our dataset traces only the mean behavior of AGN spectra, which may exhibit more or less scatter than a set of measurements tracing the instantaneous states of the spectra. The spectra were all corrected for Galactic extinction assuming a \citet{fit99} reddening law with extinction values derived from \citet{sch11}. 
 In regions of overlap among the datasets, we used measurements from COS.

\subsection{Continuum Measurements}

Although AGN continua are well characterized locally by power laws, they are generally contaminated by broad iron emission. The high density of UV \ion{Fe}{2} lines, and to a lesser extent \ion{Fe}{3} lines, coupled with the potentially large dynamical velocities of the BLR from which the flux is thought to originate, leads to a pseudo-continuum of iron emission \citep{wil85,bor92,ves01,tsu06}. This pseudo-continuum substantially complicates continuum placement and line measurement because there are few spectral regions redward of \unit{1250}{\angstrom} that are left uncontaminated. The most common method of dealing with this contamination is to fit a template of iron emission to the spectrum before subtracting it out. Because theoretical templates have thus far been unable to accurately reproduce the observed \ion{Fe}{2} emission \citep{sig03,ver04}, some authors have constructed empirical templates based on narrow-line Seyfert 1 (NLSy1) galaxies, especially I Zw 1 \citep[e.g.,][]{bor92,ves01}. By assuming that the iron continua among AGN differ by only a scale factor and a velocity dispersion, such a template can be applied to other objects. This approach has been adopted in numerous UV studies, including \citet{ves06} and \cite{she12}, and it has also been applied to studies in other wavelength regimes as redward as the near-infrared \citep[e.g.,][]{gar12}.

We adopt a similar approach in this study; the spectra for each object in our sample were fitted globally with  a power-law and an iron template. For our iron model, we used a version of the \citet{ves01} \ion{Fe}{2} template supplemented by the \citet{tsu06} template at wavelengths greater than \unit{3090}{\angstrom}. A notable shortcoming of the template is its failure to account for iron emission in the range $2770-2818$~\AA, which is coincident with the \ion{Mg}{2} line and remains poorly constrained. However, \citet{sal07} found that such an omission increases the mean measured FWHM of the \ion{Mg}{2} line by only 1.3\% compared to the theoretical template of \citet{sig03}. Therefore, we expect this template gap to contribute negligible error to our measurements. In general, we did not attempt to correct for \ion{Fe}{3} emission because it is poorly constrained \citep{gre10}, but in a few cases of obvious emission we applied the \ion{Fe}{3} templates from \citet{ves01}. The templates were broadened to the velocity width of the H$\beta$ line for each object through a convolution with a Gaussian in log-space, where the velocity scale is linear with wavelength ($d\lambda /\lambda =dv/c$). We then chose spectral regions thought to be free of emission other than from iron, to which we fit an initial power law before manually adjusting the iron template scale factor. The resulting parameters were then used as initial guesses for a five-parameter fit: two for the power-law component (slope, normalization) and three for the template (scale, velocity width, and a small wavelength shift). In contrast to \citet{bor92}, who found that the appropriate velocity width was immediately apparent and well constrained, we found that for most objects in our sample (with ${\rm FWHM_{CIV}}\gtrsim$ \unit{2000}{\kilo\meter~\reciprocal\second}) the quantity was poorly constrained because the broadening washes out the the individual features of the template. For these cases, we fixed the template to the H$\beta$ width. We do not further discuss the properties of our iron template fits because they are merely an approximate way of correcting the significant iron contamination in our other measurements, which are the focus of this paper. 

All UV continuum luminosities throughout this paper were derived from the power-law component of our pseudo-continuum fit assuming luminosity distances from a $\Lambda$CDM model. They are listed in Columns 2-4 of Table~\ref{tab:agnuvlum}, reported in the rest-frame of the object, and computed using the measured redshift.

\subsection{Emission-Line Measurements}
\subsubsection{Emission-Line Profile Fits}
The pseudo-continuum models were subtracted from the original spectra, leaving the residual emission-line spectra which we fitted with a sum of Gaussians plus a linear continuum component intended to account for small, local deviations from the global continuum fit. We do not ascribe any physical meaning to the parameters of the individual Gaussian functions; their purpose is only to reproduce the line profiles. In a few cases in which the wings of strong nearby absorption complicated the local continuum fit, we used a polynomial instead of a linear function. These fits were performed using the \texttt{MPFIT} implementation of the Levenberg-Marquardt technique \citep{mar09}. Narrow absorption features, which may be either intrinsic or intervening, were manually masked before the fit was conducted. These absorption features are easily identified in \textit{FUSE} and COS data, but they are unresolved in the \textit{IUE} data. The presence of unidentified absorption in \textit{IUE} data may introduce additional error to those measurements in this and other studies \citep[e.g.,][]{ves06}. Line properties were measured from the fits rather than directly from the data. This approach mitigates the effect of the narrow absorption lines that are prevalent in some spectra, while also allowing the separation of blended lines. For consistency, we adopted this approach even in unblended cases that did not suffer from absorption.

We did not attempt to subtract a narrow component from any of the emission lines. Though some authors argue that lines such as the \ion{C}{4} doublet contain a distinct, removable narrow component arising in the narrow-line region (NLR) \citep[e.g.][]{sul07,she12}, others suggest that the narrow component is weak and difficult to reliably remove \citep[e.g.,][]{will93}. Even \citet{sul07} note that their narrow \ion{C}{4} components remain stronger and broader than expected from narrow forbidden lines like [\ion{O}{3}] (5007 \AA). This issue is complicated by evidence that, at least in \ion{C}{4} lines, some of the flux that does not trace a gravitationally dominated region may arise from outflows \citep{wan11,den12}. To some degree, these complications are likely present in the emission profiles of all highly ionized species, and they likely vary with the luminosity and orientation of the AGN. We therefore attempt to understand how variations in line profile shape affect single-epoch mass estimates rather than adopting an unreliable subtraction of NLR emission.

Though AGN spectra feature a number of strong UV emission lines \citep{shu12}, we focus on a handful of lines that are relatively isolated and can be reliably measured. In particular, we do not address the \ion{C}{3}] $\lambda 1909$ line that \citet{gre10} and \citet{ho12} inconclusively suggested as an SMBH mass tracer because the poor quality of the \textit{IUE} data does not allow reliable separation of the line from the \ion{Al}{3} $\lambda 1857$ and \ion{Si}{3}] $\lambda 1892$ lines with which it is blended. We similarly avoid all lines that are typically blended with Ly$\alpha$, such as the strong \ion{N}{5} and \ion{Si}{4} doublets in that region.

\noindent
Details of the individual emission-line fits are as follows:
\begin{itemize}

\item \ion{C}{4} (1548.19 \AA, 1550.77 \AA;  permitted doublet [$2p(^2P)\rightarrow 2s(^2S)$]). We used up to three pairs of Gaussians, with each pair fixed with the expected velocity difference and flux ratio and assumed to have the same line width. We fit the range \unit{1500-1600}{\angstrom}. We did not fit the \ion{He}{2} $\lambda1640$/\ion{O}{3}] $\lambda 1663$ complex simultaneously with the \ion{C}{4} lines  because they are expected to contribute negligible flux shortward of \unit{1600}{\angstrom}. We assumed that the 1600 \AA~feature of unknown origin is not \ion{C}{4}.   Instead, we treated it as continuum when fitting the additional linear continuum component. This approach is unphysical and likely introduces additional scatter into our line measurements.  However, owing to the unknown source of the feature\footnote{There has been much debate regarding the origin of this feature. A portion of the flux may be due to \ion{C}{4}, and \ion{Fe}{2} $\lambda 1608$ certainly contributes to it as well. Neither of these contributions, however, seems able to account fully for the observed flux. A full treatment of this issue is beyond the scope of this paper, but useful discussions can be found in \citet{lao94}, \citet{mar96}, and \citet{fin10}. }  more complicated prescriptions are just as likely to introduce error. Compared to using the original pseudo-continuum fit alone, this prescription yields line luminosities that  are on average 6.2\% lower and FWHMs that are on average 3.5\% lower. Figures~\ref{fig:civfits} and \ref{fig:civfits2} show the 35 \ion{C}{4} emission lines and the fits to their profiles.

\begin{figure*}
\epsscale{2.1}
\plotone{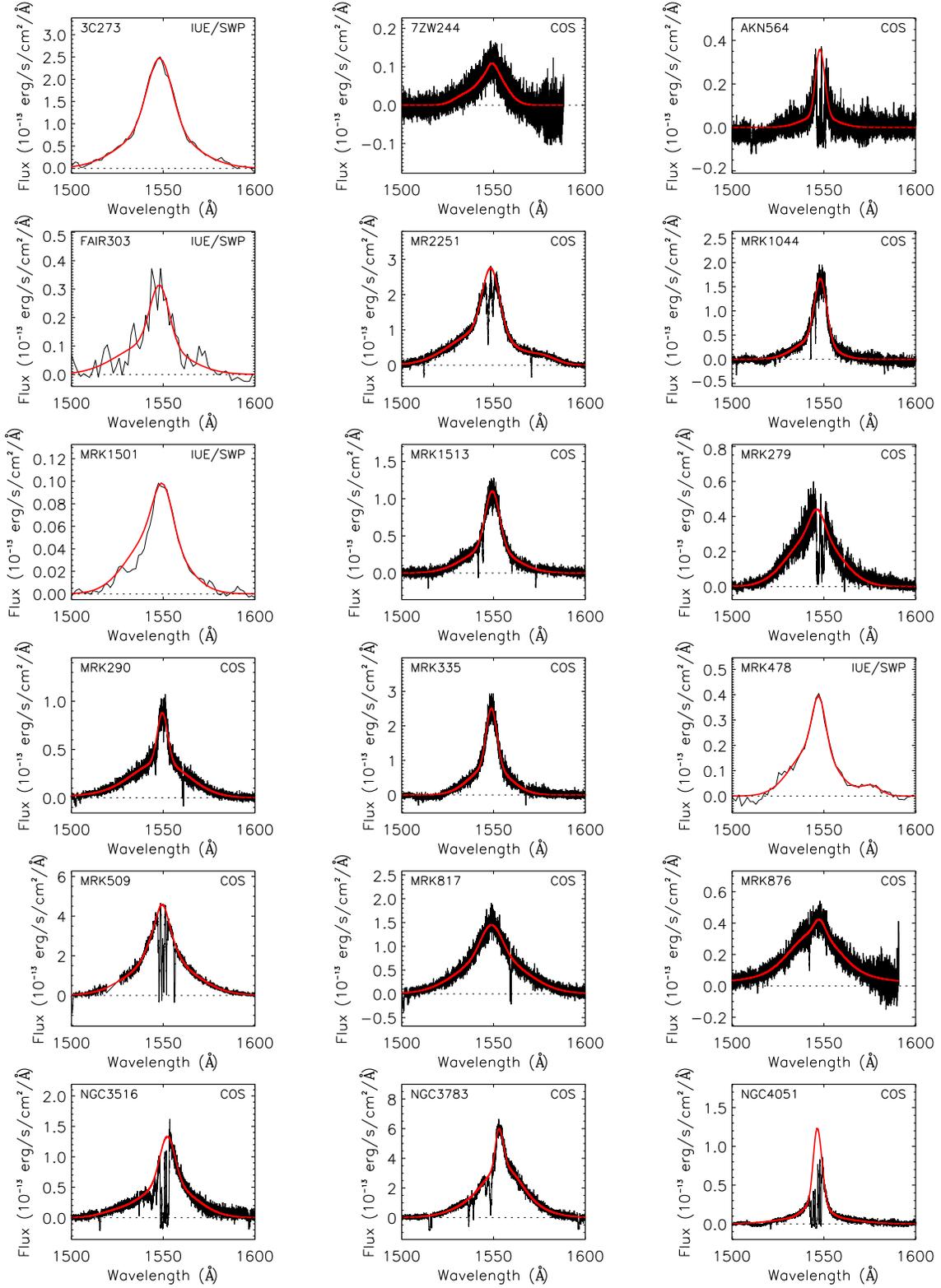}
\caption{The \ion{C}{4} ($\lambda$1548,1551) emission-line profiles. The black line is the pseudo-continuum-subtracted, emission-line residual data, and the red lines are the fits to these profiles. See Section~2.4 for details. \label{fig:civfits}}
\end{figure*}

\begin{figure*}
\epsscale{2.1}
\plotone{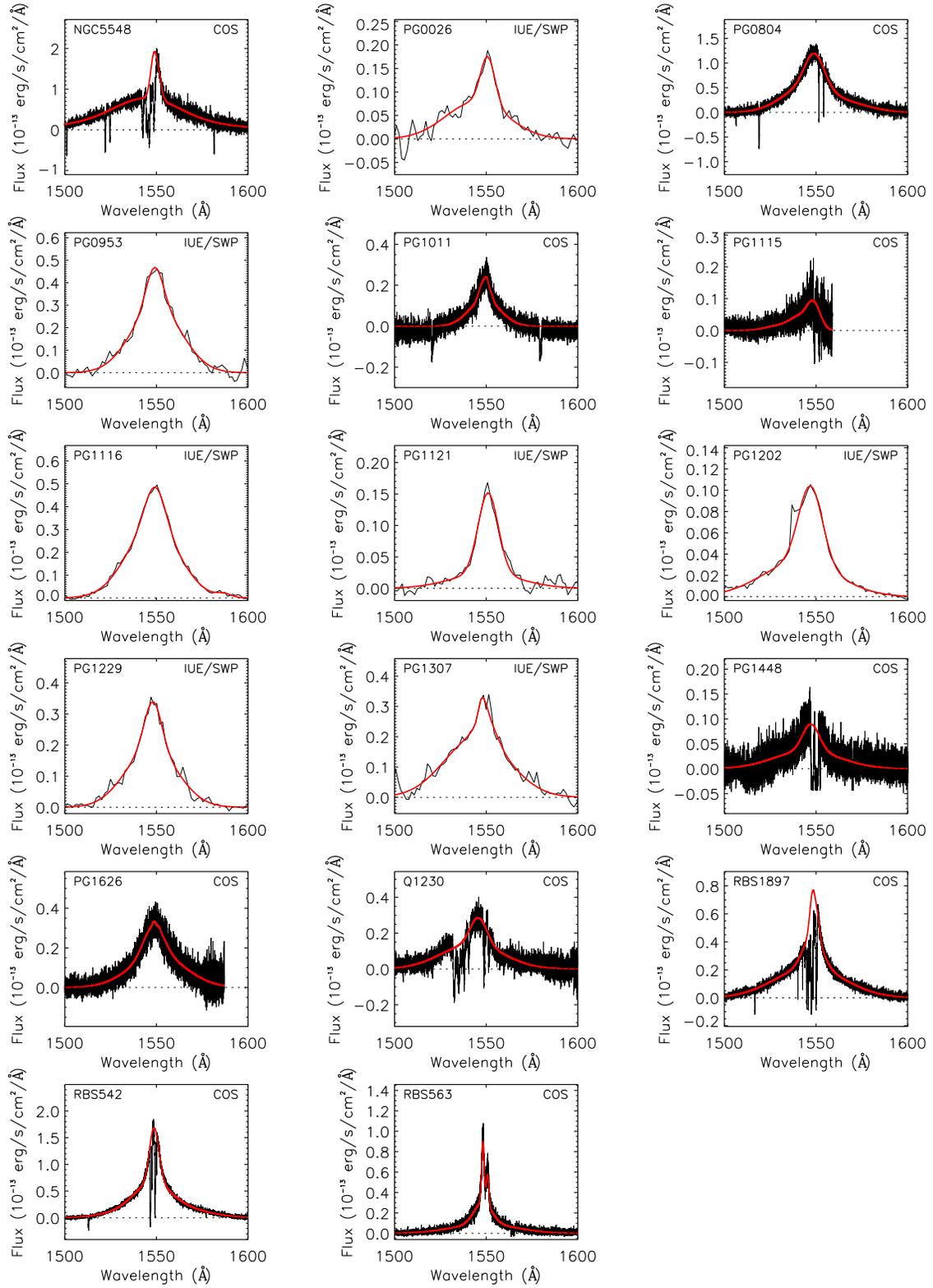}
\caption{Continuation of Figure~\ref{fig:civfits}.\label{fig:civfits2}}
\end{figure*}


\begin{figure*}
\epsscale{1.3}
\plotone{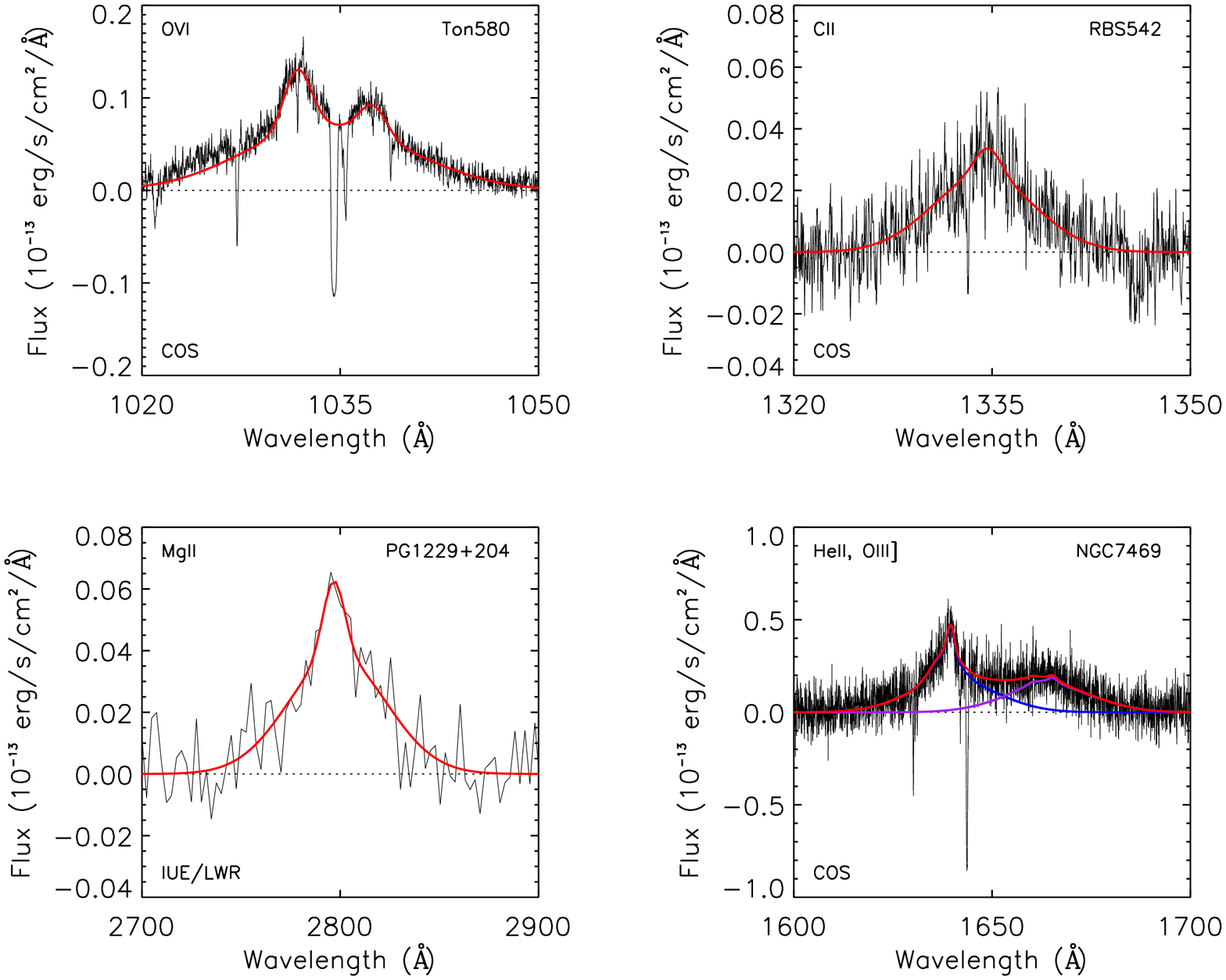}
\caption{Typical line profiles toward four AGN for \ion{O}{6} ($\lambda\lambda$1032, 1038), \ion{C}{2} ($\lambda\lambda\lambda$1334, 1336, 1336), \ion{Mg}{2} ($\lambda\lambda$2796, 2804), and the \ion{He}{2} ($\lambda$1640) and \ion{O}{3}]  ($\lambda\lambda$1661, 1666) complex. The black line is the pseudo-continuum-subtracted, emission-line residual data, and the red lines are the fits to these profiles. In the lower-right panel, the blue line and purple line are the \ion{He}{2} and \ion{O}{3}] contributions to the fit, respectively. The plotted COS data has been binned by three pixels to improve plot legibility.\label{fig:otherfits}}
\end{figure*}

\item \ion{O}{6} (1031.93 \AA, 1037.62 \AA; permitted doublet [$2p(^2P)\rightarrow 2s(^2S)$]). \ion{O}{6} is often blended with Ly$\beta$ (1025.72 \AA) and must be fitted simultaneously with that line. We fit the three lines with two sets of three Gaussians. Each set was fixed with the expected velocity differences among Ly$\beta$ and the two \ion{O}{6} lines, and the set was fixed to have the same line widths. The two \ion{O}{6} lines within a set were additionally fixed to have a flux ratio between 1:1 and 2:1, with the range in this parameter allowing for uncertain optical depth effects. We fit the range \unit{1000-1060}{\angstrom}. The top-left panel of Figure~\ref{fig:otherfits} shows a typical \ion{O}{6} profile and fit.

\item \ion{C}{2} (1334.43 \AA, 1335.66 \AA, 1335.71 \AA; three permitted lines [$2s2p^2(^2D)\rightarrow 2s^2 2p(^2P)$]). We fit the range \unit{1310-1365}{\angstrom} with three Gaussians.The top-right panel of Figure~\ref{fig:otherfits} shows a typical \ion{C}{2} profile and fit.

\item \ion{Mg}{2} (2796.35 \AA, 2803.53 \AA; permitted doublet [$3p(^2P)\rightarrow 3s(^2S)$]).   We fit the range \unit{2700-2900}{\angstrom} with up to three Gaussians. The spectral quality is inadequate to fit this line as a doublet. Note that this doublet is accessible only in the \textit{IUE} data. The bottom-left panel of Figure~\ref{fig:otherfits} shows a typical \ion{Mg}{2} profile and fit. 

\item \ion{He}{2} (1640.5 \AA; Balmer-$\alpha$ line) and \ion{O}{3}]  (1660.81 \AA, 1666.15 \AA; two semi-forbidden lines [$2s2p^3(^5S)\rightarrow 2s^2 2p^2(^3P)$]). The \ion{He}{2} and \ion{O}{3}] lines are usually heavily blended and must be fitted simultaneously. We used up to three sets of three Gaussians, with each set fixed to have the same velocity width and expected velocity difference. We additionally fixed the two \ion{O}{3}] lines within a set to have the expected optically thin flux ratio of 1:2.48. We fit the range \unit{1600-1700}{\angstrom}. It is possible that the \ion{O}{3}] lines suffer some contamination from the \ion{Al}{2} $\lambda 1670$ line. However, we are unable to detect any evidence of this line and adding additional Gaussian components at its velocity offset does not improve the fit. The bottom-right panel of Figure~\ref{fig:otherfits} shows a typical profile and fit to this complex.
\end{itemize}

\subsubsection{Emission-Line Parameter Measurement}

We characterize the widths of the emission lines using a variety of measures. We calculate the FWHM for single- and double-peaked emission lines according to the procedures described by \citet{pet04}. We similarly calculate the full-width at quarter, third, and three-quarters maximum (FWQM, FWTM, and FW3QM, respectively), and we measure the line dispersion, $\sigma_l$ (the square-root of the second moment of the profile). The dispersion is more sensitive to the wings of the profile than the FWHM, but it is also more sensitive to continuum placement, iron subtraction, and limits of integration. All line widths are corrected for instrumental resolution according to the \citet{pet04} prescription. This correction is negligible for the relatively high-resolution FUSE and COS data, but it can be significant for measurements made from IUE data. We adopt instrumental resolving powers, $R$, of 18,300 for COS, 20,000 for FUSE, 540 for the long-wavelength channel of IUE, and 260 for the short-wavelength channel of IUE. 

In an attempt to find other useful parameters for the characterization of the line profiles, we also measured the centroid relative to the \ion{Mg}{2} line (where possible), skewness ($s$, the third standardized moment), kurtosis ($k$, the fourth standardized moment), and the blueshift-asymmetry index (BAI) described by \citet{wan11}. The BAI did not strongly correlate with any other properties of the profile or AGN for any emission line, perhaps owing to scatter in our adopted systemic redshifts; we therefore do not discuss this parameter further. For the error estimates of the aforementioned properties, we adopt the maximum range in the values of a given property that is allowed by varying the fit parameters of the Gaussians independently within the parameters' $1\sigma$ error bars as returned by \texttt{MPFIT}. This range is determined numerically by generating random values of the fit parameters uniformly distributed over their $1\sigma$ errors before recalculating the line profile properties from each randomized ensemble of Gaussians. We then take the range in these recalculated properties as the error estimates for those properties. These error estimates are not true $1\sigma$ errors such as what would result from a standard Monte Carlo process. Instead, they are slightly larger, and we believe that they better represent the uncertainties present in the measurements.  The luminosities of the lines were determined from the rest-frame fluxes of the fits and measured redshifts, with errors in the fit parameters propagated forward.  The line luminosities are reported in Table~\ref{tab:agnuvlum}, and the FWHM and $\sigma_l$ measurements are reported in Table~\ref{tab:agnuvwidth}. We do not explicitly tabulate the other measured parameters discussed in this paper, but those measurements are available upon request.

\input{agnuvlumtable}


\input{agnuvwidthtable}

\begin{figure*}
\epsscale{1.}
\plotone{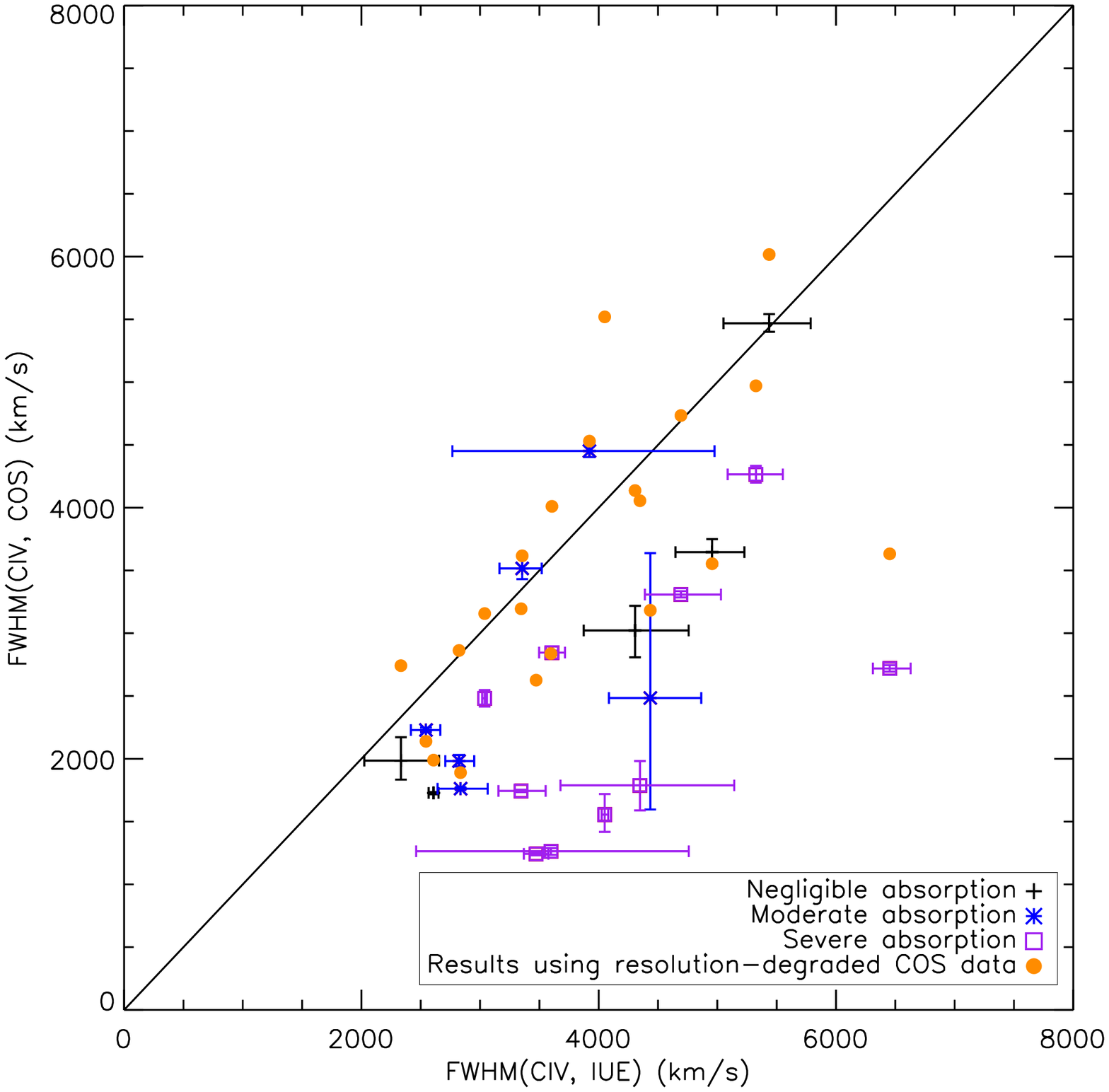}
\caption{The FWHM of the \ion{C}{4} emission line in 21 objects for which we have measurements from both COS and \textit{IUE}. The black, blue, and purple data points are the original width measurements corrected for broadening due to resolution following the \citet{pet04} prescription described in the text. The color indicates the severity of absorption contaminating the profile. The \textit{IUE} values are systematically larger. The solid orange data points are the widths of the same emission lines, measured using COS data that degraded to the resolution of IUE data. These points are consistent with a one-to-one relationship (black line), suggesting that  lower resolution is the source of the systematic offset, likely due to unresolved absorption and non-Gaussian line shapes.\label{fig:iuevscos}}
\end{figure*}

We note that \textit{IUE} line widths are systematically larger than those measured from COS data, for AGN in which we have \ion{C}{4} coverage in both datasets. This effect is illustrated in Figure~\ref{fig:iuevscos}. The black, blue, and purple data points are the original width measurements corrected for broadening due to resolution following the \citet{pet04} prescription described above. The color indicates the severity of absorption contaminating the profile, determined according to the fraction of the emission-line fitting region that was masked by absorption in the COS data. The three groups indicate lines with more than 10\% of their profile masked (``severe''), lines with 5-10\% of their profile masked (``moderate''), and lines with less than 5\% of their profile masked (``negligible''). The \textit{IUE} values are systematically larger. The difference is likely attributable to a combination of unresolved absorption, along with differences in spectral resolution.  The \textit{IUE} measurements of severely absorbed profiles have the largest systematic offset from the COS measurements, demonstrating that broad-line widths can be biased by this effect in lower resolution spectra. These absorption lines are obvious in the COS data, but most are unidentifiable in the \textit{IUE} data.  Even the profiles that do not suffer from absorption yield line widths with a systematic offset despite having been corrected for instrumental resolution as described above. This seems to be partially caused by the non-Gaussian line shapes common to these lines. Cuspy peaks are often unresolved, which can lead to a significant change in the location of the half-maximum relative to asymmetric features or inflection points in the profile.


\begin{figure*}
\epsscale{1.5}
\plotone{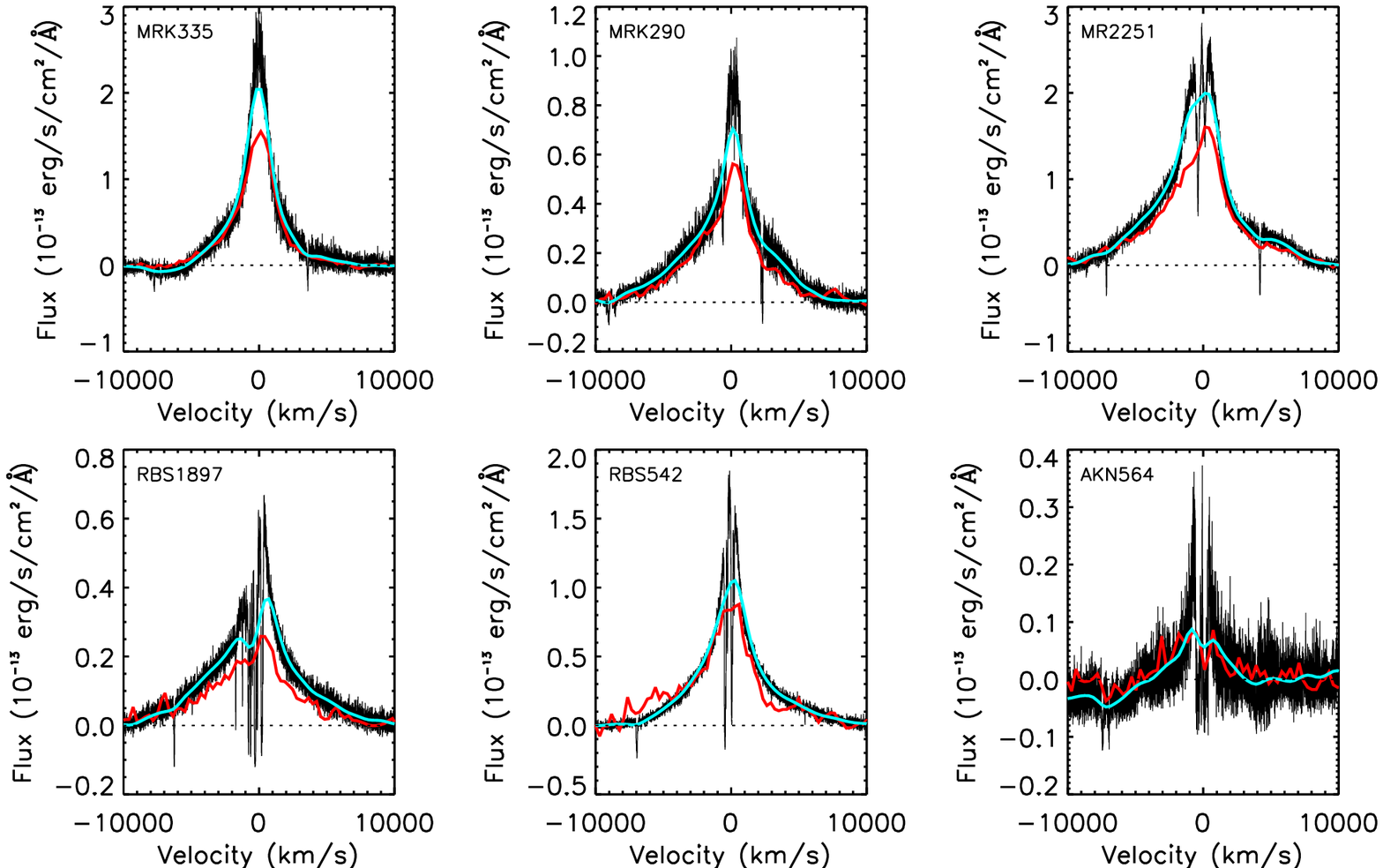}
\caption{Sample \ion{C}{4} $\lambda 1549$ spectra with COS (black), \textit{IUE} (red), and resolution-degraded COS (cyan) data, illustrating the range of effects of intrinsic AGN absorption on low-resolution spectra. The unabsorbed, roughly symmetric profile in Mrk 335 is relatively well behaved and can be simply corrected for resolution, while more extreme examples, such as those in the bottom row, behave more erratically, yielding systematically larger line-width measurements. \label{fig:iuevscosspec}}
\end{figure*}

To test if this offset is indeed caused by the difference in resolution, we degraded the COS data to the resolution of IUE. The orange points in Figure~\ref{fig:iuevscos} are the measurements made from these degraded spectra. These measurements from the resolution-degraded COS data are systematically larger than the original COS measurements, so the orange data points are shifted upward toward the one-to-one relationship (the black line in Figure~\ref{fig:iuevscos}) relative to the original data points. This result suggests that the difference in resolution is causing the systematic offset between the \textit{IUE} and COS measurements.  Figure~\ref{fig:iuevscosspec} shows several of these spectra, with \textit{IUE} data plotted in red, the COS data plotted in black, and the resolution-degraded COS data plotted in cyan. The badly absorbed profiles, such as those of RBS 542 and Akn 564, remain well-determined in the original COS data. In the degraded COS data and the IUE data, however, the absorption near the peak of the profile is unresolved, leading to a much larger line width determination. Note that this offset is independent of S/N. These results suggest that the systematic error due to resolution in width measurements made from data with resolution comparable to that of IUE  may be larger than what is expected from a simple resolution correction as most authors use. The combined result of these effects yields systematic errors as large as $\sim2000~\rm km~s^{-1}$, potentially a large source of scatter for studies that rely exclusively on data from telescopes such as \textit{IUE} or other lower-resolution instruments, such as those used by the Sloan Digital Sky Survey \citep[$R\sim 2000$;][]{ahn12}.

Although line-profile and continuum fits to the data are well suited to least-squares methods  (e.g., \texttt{MPFIT}) that account for error in the dependent variable, the same is not true of the fits to the mass-scaling relationships that we consider in this study. These relationships are characterized by large intrinsic scatter, significant measurement error in both dependent and independent variables (which may be correlated), and in some cases a small number of data points. In such cases, least-squares methods are biased toward zero \citep{akr96}. Results become unpredictable and more biased when the function has more than one independent variable \citep{kel07b}. These effects are most frequently dealt with by extending the least-squares approach  or by constructing a new ``effective chi-squared'' statistic that includes terms for these additional sources of scatter. In practice, the most common methods are the \texttt{BCES} estimator \citep[``Bivariate Correlated Errors and intrinsic Scatter,''][]{akr96} and the \texttt{FITEXY} estimator \citep{pre92}. While each of these methods has merits, they both exhibit significant bias in some cases \citep{kel07b}. The \texttt{BCES} estimator is susceptible to bias and/or inefficiency when the sample is small, when the measurement errors are comparable to the variance of the distribution, or when some measurements have errors much larger than others \citep{tre02}.

\subsection{Regression Techniques}
For these reasons, we adopt the Bayesian linear regression technique of \citet{kel07b}. This method uses a mixture of many Gaussians (assigning no physical meaning to each) to model the likelihood functions of independent variables. Uniform priors are adopted for the regression parameters to calculate their posterior probability distribution. A Markov chain Monte Carlo (MCMC) sampler is used to draw from the posterior distribution, and the best-fit parameters and their uncertainties are taken as the medians and standard deviations, respectively, of these marginalized distributions. Here, we use a Gibbs MCMC sampler. Using both simulation and astronomical data, \citet{kel07b} showed that this method is both less biased and more efficient than regressions with the \texttt{BCES} or \texttt{FITEXY} estimators. This method has the additional benefit of providing robust measurements of the intrinsic scatter in the distribution being fit, which we take as the median of the distribution of the  standard deviations of the individual MCMC realizations.

\label{sec:fitting}

\section{CORRELATIONS AMONG PROPERTIES}
\label{sec:cor}

If an emission line traces the gravitational potential of the SMBH, one would expect that the properties of its profile would correlate with the H$\beta$ parameters used to estimate the SMBH mass. We therefore computed the complete correlation matrix using the Spearman rank-order correlation coefficient, $\rho$, among all of the measured parameters, as well as the H$\beta$ parameters from the literature. This statistic's absolute value ranges from 0 to 1, where  $|P|=1$ indicates that the two variables are perfectly related by a monotonic function and  $|P|=0$ indicates no monotonic correlation. We define $P$ as the significance of a correlation coefficient's deviation from zero, with values close to zero indicating higher significance of the correlation. This value approximates the probability of the correlation arising by chance.

\citet{ves06} showed that the H$\beta$ line luminosity, $L_{\rm H\beta }$, as well as several different continuum luminosities could could act as reliable tracers of the radius, $R_{\rm BLR}$, of the broad-line region. The correlations among our measured luminosities support and extend this finding. These results can be seen in Table~\ref{tab:lumcor}, where the values above the diagonal are the rank-order coefficients and the values below the diagonal are their significances. All UV line luminosities are well correlated with $L_{\rm H\beta }$ as well as with each other.  Similarly, the monochromatic continuum luminosity at 1450~\AA~traces $L_{\rm H\beta }$. This result extends to continuum luminosities at other wavelengths, and we find no difference in magnitude of the correlation coefficients or their significances when a different wavelength is used. The power-law slope shows no significant correlation with either $L_{\rm H\beta }$ ($\rho=-0.11$; $P=0.49$) or SMBH mass ($\rho=-0.14$; $P=0.37$), which might be expected if the relation between continuum luminosity and 
$R_{\rm BLR}$ depended significantly on the wavelength of the luminosity diagnostic. 


\input{corlumtable}


\input{corwidthtable}

The correlations of various line widths with SMBH mass have been more contentious. \citet{she12}  found no correlation of $\rm (FWHM)_{CIV}$ with $\rm (FWHM)_{H\beta}$ in their sample, despite the strong correlations found by \citet{ves06} at slightly lower luminosities. The correlations among line-width measurements in our sample are less clear than the luminosity correlations.  Table~\ref{tab:widthcor} presents the correlation matrix for the FWHM measurements. All lines except for those of \ion{He}{2} and \ion{C}{2} are significantly correlated with $\rm (FWHM)_{H\beta}$, although the strengths of those correlations vary substantially. This suggests that, to some extent, these FWHMs are tracing the same gravitational potential as $\rm (FWHM)_{H\beta}$, but with substantial scatter. For comparison, $\rm (FWHM)_{H\beta}$ and the masses derived from it as described in Section~\ref{sec:mass} correlate only at the $\rho=0.5$ level. From this study alone, it is impossible to determine whether the differences among the line widths arise from contamination of the UV lines, H$\beta$ itself, or some combination of the two. Nonetheless, what correlations we do see suggest that these UV line widths may have potential utility for the construction of single-epoch mass-scaling relationships.

For all of the emission lines investigated, the other measures of line width display similarly significant correlations with $\rm (FWHM)_{H\beta}$ as do the FWHM measurements. In general, the correlations with $\rm (FWHM)_{H\beta}$ slightly improve as the width measurement moves down the line profile, except for in the case of \ion{C}{4}. Taken alone, FWTM and FWQM are generally marginally better tracers of $\rm (FWHM)_{H\beta}$ than FWHM, and in all cases, FW3QM is more poorly correlated with $\rm (FWHM)_{H\beta}$ than the other measures. This result may indicate that the core of the line profile suffers from more non-gravitational contamination than the wings, perhaps owing to the presence of NLR flux. The line dispersions, $\sigma_l$, correlate with $\rm (FWHM)_{H\beta}$ at levels similar to those of the other line-width measurements: $\rho_{\sigma,{\rm CIV}}=0.56$ ($P=0.00045$), $\rho_{\sigma,{\rm MgII}}=0.67$ ($P=0.00031$) $\rho_{\sigma,{\rm OVI}}=0.46$ ($P=0.011$), $\rho_{\sigma,{\rm HeII}}=0.47$ ($P=0.023$), $\rho_{\sigma,{\rm OIII}}=0.49$ ($P=0.028$). The \ion{C}{2} measurements again show no correlation. The larger errors in the $\sigma_l$ measurements compared to the FWHM measurements may obscure some of the behavior of that line-width diagnostic. However, as with the correlations of the UV FWHMs with $\rm (FWHM)_{H\beta}$, these correlations suggest possible utility as virial estimators.

Among the other line-profile diagnostics, we see little evidence of useful correlation with SMBH mass. The only significant correlations are the kurtosis ($\rho=-0.44$, $P=0.04$) and skewness ($\rho=0.49$, $P=0.02$) of the \ion{Mg}{2} line, though similar trends in these two parameters are seen in the other lines at low significance ($P\sim 0.1$). Notably, we do not see a significant correlation of mass with blueshift of \ion{C}{4} relative to \ion{Mg}{2}, as might be expected from the correlation of \ion{C}{4}-based mass residuals with that line's blueshift relative to the Balmer lines found by \citet{she12}. We have only thirteen objects with which to test this particular dependence.

\begin{figure*}
\epsscale{1.75}
\plotone{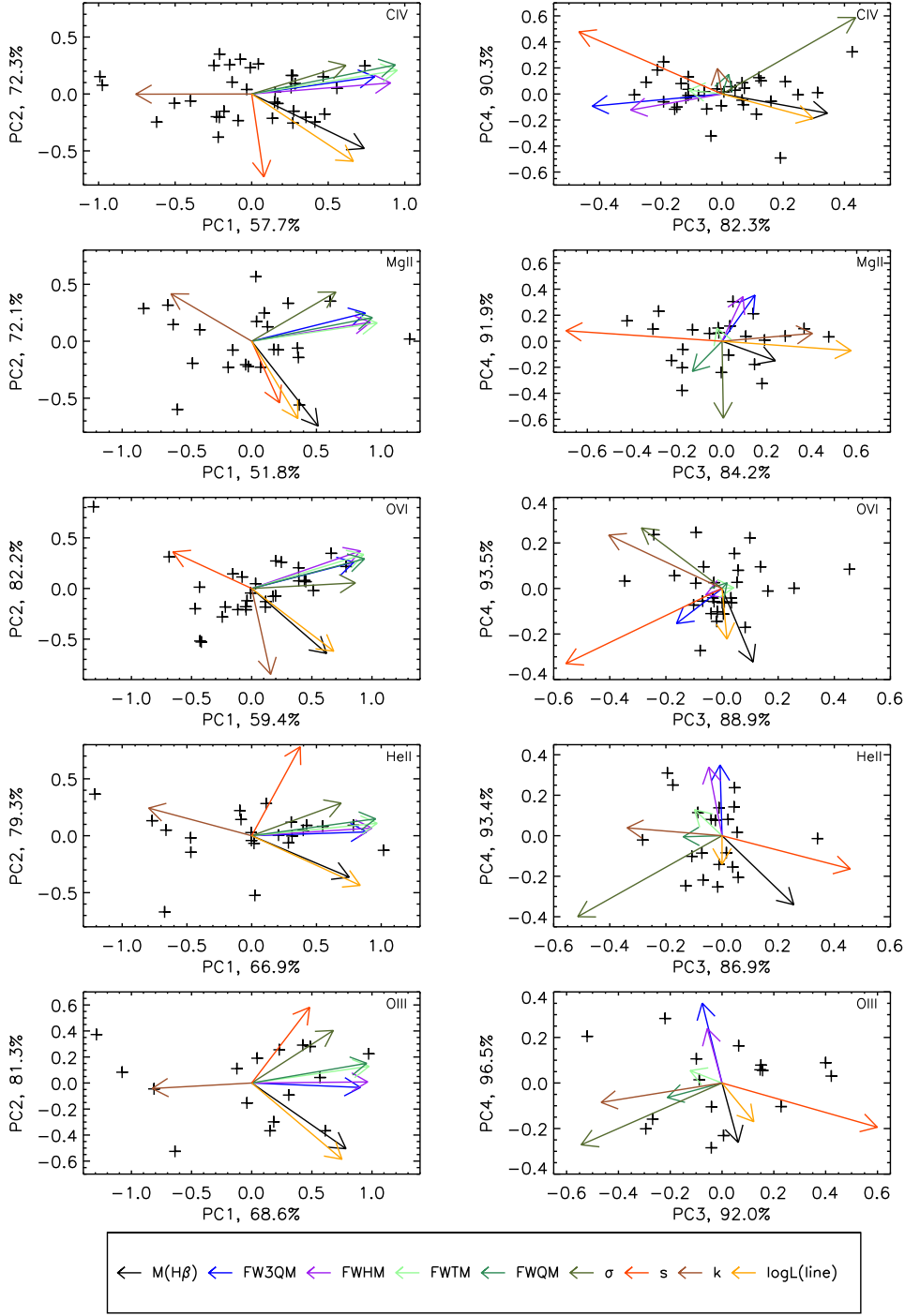}
\caption{Projections of data (crosses) and variables (arrows) onto the first four principal components from nine-variable ($M_{BH}$, FWQM, FWTM, FWHM, FW3QM, $\sigma_l$, $k$, $s$, $\log L_{\rm line}$) principal component analyses for five emission lines. The cumulative percentage of the variance of the total data set accounted for by a given component is given in each axis label. Note the strong correlation of mass with line-width diagnostics and line luminosities in the first two principal components of each line. See Section~\ref{sec:cor} for more details. \label{fig:pca}}
\end{figure*}

The potentially complicated interrelationship of these line diagnostics and their possible connection to SMBH mass is not fully captured by a series of monovariate analyses. Complicated sets of variables may have covariances that mask their interrelationship unless correlation analyses are conducted in a multivariate manner.  We therefore illustrate the correlations among these variables using a principal component analysis (PCA). Using a correlation matrix, a number of correlated variables are transformed into a smaller number of uncorrelated vectors (the principal components) by decomposing the normalized correlation matrix into a set of eigenvectors whose eigenvalues reflect the ability of that principal component to reproduce the variance of the dataset. We performed a variety of PCAs using various subsets of the measured emission-line parameters along with the SMBH masses. Perhaps the easiest way to visualize the results of such an analysis is to look at how the initial parameters project onto the principal components. For illustrative purposes, we present a simplified set of PCAs in Figure~\ref{fig:pca}, in which each set of line measurements is described with just nine variables ($M_{BH}$, FWQM, FWTM, FWHM, FW3QM, $\sigma_l$, $k$, $s$, $\log L$). For each emission line, we plot the projections of the data (cross data points) and the variables (arrows) onto the first four principal components (the axes of the plots). The length of an arrow's component in a particular axis thus indicates its strength in that principal component, so a variable with its arrow strongly projected onto a particular principal component has a large amount of its variance explained by that principal component. Each axis label additionally contains the cumulative percentage of the total variance of the dataset that is accounted for by that principal component. The parameters that most affect $M_{BH}$ are those vectors with with large components parallel or antiparallel to the $M_{BH}$ vector.


A PCA places as much of the total variance as possible in each subsequent component. Thus, if most of a dataset's variance can be accounted for by the first several principal components, then some of the variables in that dataset are redundant for explaining the variance. We can immediately see that the variables plotted here carry a great deal of redundant information, with over 90\% of each dataset's variance preserved in the first four components. The near-coincidence of the vectors for the various width measurements in Figure~\ref{fig:pca} indicate that they are all highly correlated and likely all trace $M_{BH}$ to similar degrees.  The \ion{Mg}{2} plots are perhaps the clearest, with the first two principal components dominated by the line luminosity and width measurements. Mass is substantially projected onto both of these components, indicating that line widths and luminosities do indeed trace black hole mass, and that these two measures carry different but overlapping information. We can contrast these results with those of \ion{He}{2} or \ion{O}{3}], where the projections of line luminosity and width move closer together.  There is evidently more overlap in the information conveyed by these diagnostics, and our line widths may just be tracing changes in line luminosity.  The third and fourth principal components are substantially more confused. In the fourth component of \ion{Mg}{2}, we see a substantial projection of mass onto the other, anti-correlated line-shape parameters ($k$, $s$). This may indicate that a third (shape) parameter could be useful in improving single-epoch mass estimates as a negative correction to the masses derived from line widths and luminosities alone. In \ion{C}{4}, however, the kurtosis dependence is weaker and reversed. This difference may indicate that the fourth principal component of \ion{Mg}{2} is driven by the lower resolution of \textit{IUE} data. Nonetheless, the overall picture presented by these PCAs is a confirmation of the monovariate correlation analysis: line width and luminosity strongly trace SMBH mass, and other line-shape diagnostics may act as a third parameter.

To summarize the overall picture presented in these correlation analyses, we find that the mass of the SMBH is separately correlated with line width and luminosity measures when using the \ion{C}{4} and \ion{Mg}{2} emission lines and, to a lesser degree, the \ion{O}{6}, \ion{He}{2}, and \ion{O}{3}] emission lines. While there are suggestions of dependence of SMBH mass on a third line-shape parameter, such as kurtosis or skewness, we cannot yet arrive at any strong conclusions.   The \ion{C}{2} lines do not show any significant correlations, and we therefore proceed no further in trying to construct single-epoch mass-scaling relationships from them. The \ion{C}{2} lines are typically the least luminous of the emission lines studied here, so this may simply reflect the smaller sample size and larger errors associated with the measurements. The difference could also represent a true physical difference in where the gas resides, but this explanation seems somewhat unlikely given the similar ionization potentials of \ion{C}{2} and \ion{Mg}{2}. A larger sample of high S/N \ion{C}{2} spectra would be necessary to resolve these issues.

\section{MASS-SCALING RELATIONSHIPS OF UV EMISSION LINES}

In light of the evidence described in Section~\ref{sec:cor} that FUV emission-line widths do, in fact, trace SMBH mass, we attempted to obtain single-epoch mass-scaling relationships with all of the emission lines except \ion{C}{2}. These scaling relationships will serve as a check on the consistency of other calibrations in the literature. In practice, this amounts to fitting the measured line parameters to the known masses with a function of the form,
\begin{eqnarray}
\log \left ( \frac{M_{\rm BH}}{M_{\astrosun} }\right ) = \alpha+\beta\log&\left ( \frac{L}{10^{44}~{\rm erg~s^{-1}}}\right )\nonumber\\
&+\gamma\log\left ( \frac{\Delta v}{{\rm km~s^{-1}}}\right ),
\end{eqnarray}
where $L$ is some measure of luminosity and $\Delta v$ is some measure of line width.  Typically, it is assumed that $\gamma=2$, and most authors fix that parameter in the fit. Some authors, such as \citet{she12}, advocate relaxing this virial condition to account for possible covariance among the parameters and other systematic effects. Owing to our relatively small sample, the errors on the fit parameters grow substantially larger with three free parameters instead of two.  When leaving $\gamma$ free, we always obtain results consistent with $\gamma=2$, and we leave this parameter fixed for the remainder of this discussion. This approach is in line with most of the other calibrations in the literature, including \citet{ves06}, \citet{ves09}, and \citet{ass11}.

\begin{figure*}
\epsscale{1.95}
\plotone{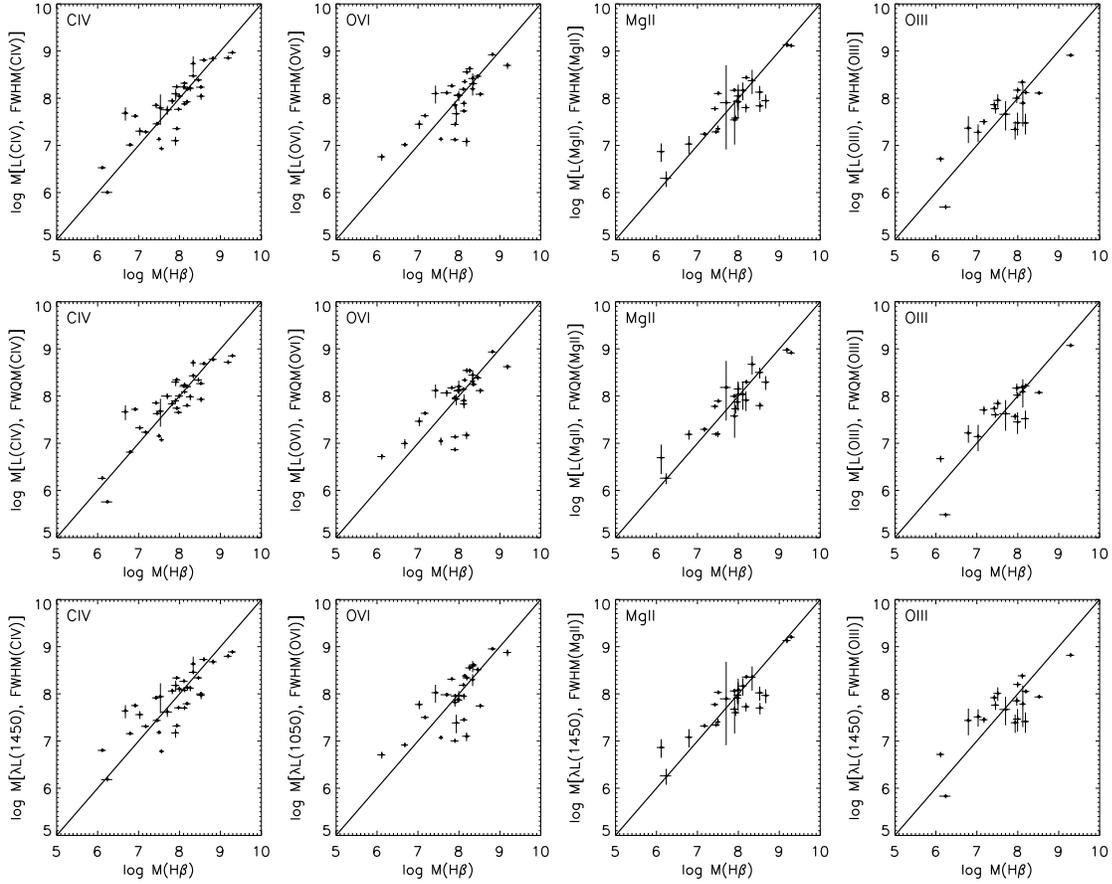}
\caption{SMBH masses resulting from the two-parameter fits in Table~\ref{tab:fitparam} plotted against the reference masses based on H$\beta$ measurements as listed in Table~\ref{tab:agn}. The reference masses are either single-epoch estimates based on the relationships from \citet{ves06} or reverberation mapping results from the literature. See Section~2.1 for more details. The solid line is a one-to-one relationship. \label{fig:massfits}}
\end{figure*}
It is less obvious how one should deal with the slope, $\beta$, of the luminosity dependence which simple photoionization models suggest should be $\beta\sim 0.5$ \citep[e.g.,][and references therein]{wan85,net93}. Optical calibrations typically fix this parameter using the results of separate studies of the $R_{\rm BLR}-L$ relationship. The \citet{ves06} H$\beta$ calibrations, for example, are fixed at slopes consistent with the observational results of \citet{kas05} and \citet{ben06}. For their \ion{C}{4} relationships, they fix the slope of the continuum luminosities, $\lambda L_{1350}$ and $\lambda L_{1450}$, at $\beta=0.53$. This value is chosen to match their adopted $R_{\rm BLR}({\rm H\beta})-L_{1350}$ relationship, which \citet{ves06} argue is consistent with the somewhat uncertain slope of $\beta=0.61\pm 0.05$ derived by \citet{pet05} for a small sample of AGN with \ion{C}{4} reverberation mapping results. More recently, \citet{kas07} found $\beta=0.52\pm0.04$, consistent with the aforementioned studies, using a sample of \ion{C}{4} measurements from six AGN at higher redshift ($2.2\le z \le3.2$) and higher luminosity ($46.8\le \log\lambda L_{\lambda}(5100)\le 47.5$) than past work. There is no particular reason that these results should be valid for other continuum luminosities and emission lines. However, no reliable reverberation mapping has been performed on a sufficiently large sample of AGN to estimate the slope for lines other than the Balmer lines and, to a lesser extent, the \ion{C}{4} line. We therefore perform two sets of fits, one in which we leave $\beta$ as a free parameter and one in which we fix $\beta=0.53$. The latter approach assumes constant electron density throughout the BLR gas responsible for the different lines as well as constant spectral shape across AGN. These are likely imperfect assumptions, and results deriving from them should be viewed skeptically until they can be confirmed with more extensive reverberation mapping results than are presently available.  In most cases our fits in which $\beta$ is left free are consistent with both $\beta=0.53$ and $\beta=0.5$ as one would expect in a simple, photoionized BLR case.


\input{fitparamtable}

The results of these fits are presented in Table~\ref{tab:fitparam}. We report fits based on the line luminosities used with the FWHM, FWQM, and $\sigma_l$, as well as fits using continuum luminosities with FWHM.  Figure~\ref{fig:massfits} shows these cases for the fits with two free parameters. We have not reported the \ion{He}{2} fits because they display only a weak correlation with mass and a large scatter of $\sim 0.84$ dex. The weak observed correlation is entirely dominated by the luminosity, with the line widths insignificantly reducing the scatter. If $\gamma$ is left free in these fits, we find $\gamma\sim 0.15$, usually consistent with zero. It is strange that we see so little dependence on the \ion{He}{2} line widths, as the flux of this line is thought to trace virial motion \citep{pet99}. This may indicate a much stronger NLR flux contribution in this line compared to the others. If that were true, one would expect to observe correlations of the mass residuals with line shape; we see no such correlation with any of the moments of the line profile or any of the alternate width measurements. Alternately, the \ion{He}{2} measurements may simply be contaminated due to blending with the nearby flux from \ion{Fe}{2} and/or \ion{O}{3}] lines despite our attempts to separate these features.

The mass-scaling relationships derived from the other lines, however, are well correlated with the H$\beta$ masses, although the relationships are substantially less certain for the \ion{O}{3}] $\lambda 1664$ line. Because we have a small sample of objects with a limited range of masses and luminosities for \ion{O}{3}], these results are merely suggestive of its potential as a mass estimator rather than useful calibrations.  The scatter of the masses derived from any two-parameter fit is $\sim 0.4$ relative to the reference masses, comparable to the intrinsic scatter of the single-epoch, H$\beta$ mass-scaling relationships relative to the reverberation mapping results upon which they are based. We have compared these results to fits using the luminosity variables  alone. In all cases, the scatter is larger by $0.2$ dex or more, indicating that the line widths are essential to these relationships and the correlation is not driven solely by luminosity. This result differs from the finding of \citet{fin10} and \citet{cro11}, who found that \ion{C}{4} line widths contribute no meaningful information to the virial relationships. Those two studies used a much larger sample of AGN at a variety of luminosities and redshifts than this study, but their spectral resolution and S/N was typically much lower than the COS data used here. As \citet{ves06} found for H$\beta$, estimates using the line luminosities exhibit lower scatter than those using continuum luminosities. As expected from the analysis in Section~\ref{sec:cor}, line-width measurements that emphasize the wings of the line profiles perform slightly better than FWHM measurements. The strong correlation of the \ion{C}{4} relationship with the reference masses supports the idea that masses derived from  \ion{C}{4} line widths are consistent with those found using H$\beta$ \citep{ves06,ass11}. Similarly, the \ion{Mg}{2} relationships confirm the reliability of that line as a mass estimator, as numerous other authors have found \citep[e.g.,][]{mcl02,kol06,mcg08,ves09}. For \ion{C}{4}, we repeated the fits with only the COS measurements. This smaller sample yields results consistent with the sample as a whole.


\begin{figure*}
\epsscale{1.55}
\plotone{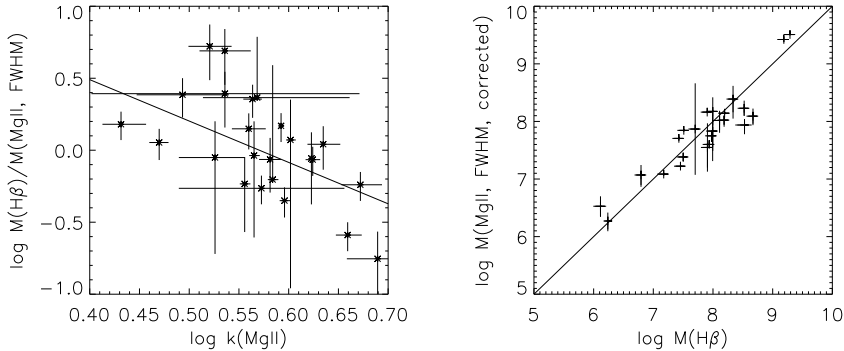}
\caption{The dependence of our \ion{Mg}{2}-based mass estimates on the kurtosis, $k$, of the line. The left panel plots the kurtosis versus the logarithm of the ratio of the H$\beta$-derived masses to the masses calculated from the fit to \ion{Mg}{2} luminosity and FWHM. The solid line is the best linear fit, $\log \left ( M_{\rm H\beta}/M_{\rm MgII} \right )=1.64-2.87\log k$.  The right panel shows the $H\beta$ masses plotted against the \ion{Mg}{2} masses corrected with the fit to kurtosis. This plot is a corrected version of the third panel in the first row of Figure~\ref{fig:massfits}. The solid line is the one-to-one relationship.\label{fig:kurtcorrect}}
\end{figure*}
In an attempt to further reduce the scatter in these relationships, we performed an additional correlation analysis on the residuals of these fits. This analysis was motivated by \citet{den12}, who found that \ion{C}{4} kurtosis as traced by the ${\rm FWHM}/\sigma_l$ ratio correlates with the residuals of masses estimated using the calibrations of \citet{ves06}. We see this dependence (Figure~\ref{fig:kurtcorrect}) only in the case of \ion{Mg}{2}, for the two-parameter FWHM fit, $\rho=-0.65$, $P=0.00053$. The best linear fit to this dependence is $\log \left ( M_{\rm H\beta}/M_{\rm MgII}\right )=(1.64\pm0.21)-(2.87\pm 0.36)\log k$.  Applying this correction to the mass estimates reduces the scatter in the distribution to 0.28 dex from 0.32 dex. We obtain a consistent result by performing a four-parameter fit (an offset and the power-law slopes of luminosity, FWHM, and kurtosis with reference mass)  for the mass-scaling relationship. We only see this dependence in the one line observed only with \textit{IUE}. It is therefore possible that we are simply observing a resolution or S/N effect. The sample used by \citet{den12} is also primarily low resolution data, with the bulk of their measurements coming from \textit{IUE} \citep[$R\sim 200-600$;][]{hol82, cas83} or SDSS \citep[$R\sim 2000$;][]{ahn12}, so it could be subject to similar effects. It is also curious that the kurtosis slope is negative. Na\"{\i}vely, one might expect that an increasingly strong NLR component would deflate FWHM measurements while increasing kurtosis, thereby giving a positive slope of $\log k$ with reference mass.

The other correlations with the mass residuals are insignificant. In particular, we detect no dependence on line blueshift relative to \ion{Mg}{2} as might be expected from the blueshift correlations found by \citet{she12}. Owing to our lack of optical continuum data, we cannot directly test the conclusion of \citet{ass11} that the mass residuals correlate with the ratio of the UV-to-optical continuum luminosity, which they attribute to extinction, host starlight contamination, or non-universal AGN spectral energy distributions. We do not see any correlation of the mass residuals with the UV continuum slope. 

It is worth reiterating that the results in this section depend on the uncertainties inherent to the mass estimates upon which we have based this analysis. These are several steps removed from direct SMBH mass measurements.  Relative to the $M-\sigma$ relation, reverberation-based mass estimates are uncertain by about a factor of three \citep{onk04}. The single-epoch H$\beta$ calibrations, in turn, have a scatter of about 0.4 dex relative to the reverberation results \citep{ves06}. The single-epoch H$\beta$ mass estimates also have their own systematic difficulties, especially those resulting from starlight contamination. This effect is less important in UV lines, so it is possible that the bulk of the scatter in our results represents the range of SMBH masses against which we are calibrating rather than the estimators that we are calibrating themselves. These limitations cannot be overcome until a better sample of reverberation mapping results are available for a variety of emission lines and a variety of AGN. We emphasize that all single-epoch calibrations are statistical in nature: they yield consistent results on average, but they may be in error by up to an order of magnitude for any particular object \citep{ves06}.

\section{SUMMARY AND CONCLUSIONS}

We have conducted an \HST/COS study of UV emission lines with the goal of better understanding the systematics of single-epoch SMBH mass-scaling relationships. Using data with higher resolution and higher S/N than previous studies of these estimators, we were able to characterize UV emission-line profiles in great detail. Although we are limited by our somewhat small sample of nearby AGN, we can come to several conclusions about single-epoch mass estimates that hold in the low-luminosity AGN regime:

\begin{enumerate}

\item Through a comparison of COS and \textit{IUE} measurements, we find that the use of low-resolution spectra has a significant impact on line diagnostics such as the FWHM, which are overestimated owing to unresolved absorption and other effects of lower resolution. This may significantly affect single-epoch mass estimates. A larger sample of objects observed at both higher resolution (such as with COS) and at lower resolution (such as with \textit{IUE}) are needed to better characterize these effects.  The sample used in this study was selected on the basis of the availability of optical mass estimates.

\item At the 1450~\AA\ luminosities probed by our sample, $\lambda L_{\lambda} \leq 10^{45}$ erg~s$^{-1}$, we confirm the utility of \ion{C}{4} $\lambda 1549$ luminosities and line widths as single-epoch mass estimators, and we show that mass-scaling relationships based this sample of new COS data are consistent with past callibrations.   We confirm the results of \citet{ves06} and \citet{ass11}, which were disputed by \citet{she12}, and we do not detect the dependence of the residuals of \ion{C}{4}-based mass estimates on kurtosis seen by \citet{den12}.   Although we cannot directly test the dependence on the blueshift of \ion{C}{4} relative to H$\beta$ \citep{she12}, we do not see any dependence on the blueshift relative to \ion{Mg}{2} within the small sample of 13 objects for which we have measurements of both lines.

\item We confirm that masses estimated from \ion{Mg}{2} $\lambda 2798$ are consistent with those based on the Balmer lines. These are the only estimates with residuals significantly correlated with kurtosis. As these measurements are the only ones taken exclusively from \textit{IUE} data, it is possible that this is an effect of low spectral resolution on measured line widths.

\item We conclude that \ion{O}{6} $\lambda\lambda 1032, 1038$ line widths can be used to trace SMBH mass, albeit with larger scatter than \ion{C}{4}- or \ion{Mg}{2}-based estimates. The additional scatter could be partially due to blending with Ly$\beta$ or systematics of the wider separation in velocity of the \ion{O}{6} doublet compared to other lines. We find evidence that \ion{O}{3}] $\lambda 1664$ may also be a viable tracer, but we see no correlation of  \ion{C}{2} $\lambda 1335$ or \ion{He}{2} $\lambda 1640$ with SMBH mass.

\end{enumerate}

A complete understanding of the systematics of single-epoch SMBH mass estimators will require the careful use of high-resolution and high-S/N spectra with a large sample of objects. Ideally, such work should be done with a larger sample of reverberation-mapped objects than is currently available, but in lieu of such measurements, progress can be made through the comparison of single-epoch optical and UV observations. The present study begins this work, but it is limited by the size of its sample and the range of luminosities. More robust results could be obtained by expanding the sample with more optical mass estimates that overlap with existing UV data, with other UV datasets, or with near-simultaneous observations of UV and optical emission lines. Single-epoch mass estimates are an imperfect method of characterizing active SMBH masses, but they will be an increasingly important tool for studies of galaxy evolution. As we have discussed here, the FUV emission lines that are necessary for high-redshift observations show great promise as reliable estimators of SMBH mass.   With its high throughput and superior spectral resolution, HST/COS can provide a new standard for the strong UV emission lines in low-redshift, bright Seyferts and quasars.   For the COS moderate-resoution gratings (G130M and G160M), the recommended lines  include \CIV\ (out to $z \approx 0.15$) and \OVI\  (out to $z \approx 0.71$).   Our future efforts will involve HST/COS archival surveys, exploring multiple-epoch data with COS, comparison to SDSS quasars, and searches for dependence  of these
line profiles on Seyfert type and radio fluxes.
\acknowledgments

We thank Charles Danforth and Matthew Stevans for providing coadditions of some of the COS and FUSE datasets and for helpful conversations.  Marianne Vestergaard and  T. A. A. Sigut graciously provided their \ion{Fe}{2} templates along with helpful comments. This work was supported by STScI grant HST-AR-12825.01-A and COS-support grant NNX08-AC14G.

{\it Facilities:} \facility{FUSE}, \facility{HST (COS)}, \facility{IUE}.

\bibliographystyle{hapj}
\bibliography{apj-jour,agnpaperbib}
\end{document}

%% file: agntable-2.tex

\begin{deluxetable}{lrrcccrrrrrc}
\tabletypesize{\tiny}
\tablecaption{AGN Properties\tablenotemark{a}  \label{tab:agn}}
\tablecolumns{12}
\tablewidth{0pt}
\tablehead{
    \colhead{Object} & \colhead{$z_{\rm QSO}$} & \colhead{$z_{\rm UV}$} & \colhead{AGN} &   \colhead{$\rm FWHM_{\rm H\beta}$} & 
    \colhead{Ref.\tablenotemark{b}} & \colhead{$\log\left [L_{\rm H\beta} \right ]$} & \colhead{Ref.\tablenotemark{b}} &
    \colhead{$\log (M/M_{\astrosun})$} & \colhead{{$\log (M/M_{\astrosun})$}} & \colhead{Ref.\tablenotemark{b}} & \colhead{UV Data\tablenotemark{c}} \\
                &        &     &  Type  &  \colhead{$({\rm km~s^{-1}})$} &    &  \colhead{$({\rm erg~s^{-1}})$}     &                &
    \colhead{(H$\beta$, SE)}           & \colhead{(rms)}                              &             &   \\
    \colhead{(1)}     & \colhead{(2)} & \colhead{(3)}& \colhead{(4)}& \colhead{(5)}& \colhead{(6)}& \colhead{(7)}& \colhead{(8)}& \colhead{(9)} 
          & \colhead{(10)}  & \colhead{(11)}   & \colhead{(12)}   \
}
\startdata
     NGC4395    & 0.001064 & 0.00108   & Sy 1.8   &   833 &  1 & $37.57\pm0.084$ &  1 & $3.722^{+0.095}_{-0.121}$ & $5.556^{+0.116}_{-0.158}$ & 9 & FABSL \\
     NGC4051    & 0.002336 & 0.00357   & Sy 1.0   &  1034 &  7 & $ $ &  & $$ & $6.238^{+0.120}_{-0.155}$ & 7 &  FASL \\
     NGC3516    & 0.008836 & 0.00716   & Sy 1.5   &  5175 &  7 & $ $ &  & $$ & $7.501^{+0.037}_{-0.062}$ & 7 &  FABS \\
     NGC3783    & 0.009730 & 0.00972   & Sy 1.5   &  3555 &  1 & $41.50\pm0.045$ &  1 & $7.455^{+0.087}_{-0.109}$ & $7.474^{+0.072}_{-0.087}$ & 5 & FABSL \\
     Mrk1044      & 0.016451 & 0.01608   & NL Sy 1 &  1280 & 10 & $43.34\pm0.044$\tablenotemark{d} & 3 & $6.794^{+0.083}_{-0.103}$\tablenotemark{d} & $$ &  &  FASL \\
     NGC7469    & 0.016317 & 0.01643   & Sy 1       &  2639 &  1 & $42.00\pm0.045$ &  1 & $7.511^{+0.087}_{-0.109}$ & $7.086^{+0.047}_{-0.053}$ & 5 &  FASL \\
     NGC5548    & 0.017175 & 0.01690   & Sy 1.5    &  5821 &  1 & $41.53\pm0.045$ &  1 & $7.903^{+0.087}_{-0.109}$ & $7.827^{+0.017}_{-0.017}$ & 5 & FABSL \\
      Akn564       & 0.024684 & 0.02490   & Sy 1.8     &  1283 &  1 & $40.77\pm0.049$ &  1 & $6.115^{+0.087}_{-0.110}$ & $$ & &  FASL \\
      Mrk335       & 0.025785  & 0.02590  & Sy 1        &  1840 &  1 & $41.96\pm0.045$ &  1 & $7.174^{+0.087}_{-0.109}$ & $7.152^{+0.101}_{-0.131}$ & 5 & FABSL \\
      Mrk595       & 0.026982  & 0.02673  & Sy 1.5    &  2352 & 11 & $41.52\pm0.044$ & 11 & $7.111^{+0.087}_{-0.108}$ & $$ & & FABSL \\
      Mrk279       & 0.030451  & 0.03011  & Sy 1        &  5410 &  1 & $42.08\pm0.047$ &  1 & $8.186^{+0.087}_{-0.109}$ & $7.543^{+0.102}_{-0.133}$ & 5 & FABSL \\
      Mrk290       & 0.029577  & 0.03023  & Sy 1.5    &  5200 &  1 & $41.74\pm0.045$ &  1 & $7.936^{+0.087}_{-0.109}$ & $7.385^{+0.062}_{-0.072}$ & 7 & FABSL \\
      Mrk817       & 0.031455  & 0.03115  & Sy 1.5    &  4656 &  1 & $42.17\pm0.049$ &  1 & $8.110^{+0.087}_{-0.110}$ & $7.694^{+0.063}_{-0.074}$ & 5 & FABSL \\
      Mrk509       & 0.034397  & 0.03364  & Sy 1.5    &  3423 &  1 & $42.72\pm0.044$ &  1 & $8.195^{+0.087}_{-0.109}$ & $8.155^{+0.035}_{-0.038}$ & 5 & FABSL \\
  Fairall303      &  0.040008  & 0.03962  &  NL Sy1 & 1450 &  6 & $41.50\pm0.044$ &  6 & $6.678^{+0.076}_{-0.092}$ & $$ &  & FABSL \\
  PG1011-040 &  0.058314  & 0.05791  &  Sy 1.2   & 2010 &  1 & $41.61\pm0.045$ &  1 & $7.031^{+0.087}_{-0.109}$ & $$ &  & FABSL \\
     Mrk1513     &   0.062977 & 0.06204  &  Sy 1.5   & 2899 &  1 & $42.60\pm0.045$ &  1 & $7.976^{+0.087}_{-0.109}$ & $8.660^{+0.049}_{-0.056}$ & 5 & FABSL \\
  PG1229+204 &  0.063010 & 0.06336  &  Sy 1.0   & 3496 &  1 & $42.38\pm0.045$ &  1 & $7.996^{+0.087}_{-0.109}$ & $7.865^{+0.171}_{-0.285}$ & 5 &   ASL \\
  MR2251-178 &  0.063980 & 0.06378  &  Sy 1.5   & 6805 &  1 & $42.30\pm0.068$ &  1 & $8.522^{+0.091}_{-0.115}$ & $$ &  &   FAB \\
  PG1448+273 &  0.065       & 0.06397   &  NL Sy1  & 1330 &  6 & $42.00\pm0.044$ &  6 & $6.918^{+0.080}_{-0.099}$ & $$ & &  FASL \\
      RBS563      &  0.069        & 0.06858  & Sy 1.5     & 2410 &  2 & $42.20\pm0.044$ &  2 & $7.560^{+0.051}_{-0.058}$ & $$ & & FABSL \\
      Mrk478       &   0.079055 & 0.07735  &  Sy 1       & 1630 &  6 & $42.53\pm0.044$ &  6 & $7.428^{+0.082}_{-0.100}$ & $$ & &  FABS \\
     Ton1187     &   0.078882 & 0.07881  &  Sy 1.2    & 2980 &  6 & $42.45\pm0.044$ &  6 & $7.902^{+0.066}_{-0.078}$ & $$ & & FABSL \\
  PG1351+640 &  0.088200 & 0.08822  &  Sy 1.5   & 5571 &  1 & $42.81\pm0.044$ &  1 & $8.670^{+0.087}_{-0.109}$ & $$ & & FABSL \\
     Mrk1501      &   0.089338 & 0.08915  & Sy 1.2$^*$    & 4891 &  1 & $42.76\pm0.045$ &  1 & $8.528^{+0.087}_{-0.109}$ & $8.265^{+0.059}_{-0.069}$ & 8 &  FABS \\
  PG1411+442 &  0.089600 & 0.08968  &  Sy 1       & 2612 &  1 & $42.78\pm0.045$ &  1 & $7.994^{+0.087}_{-0.109}$ & $8.646^{+0.124}_{-0.174}$ & 5 &  FABS \\
  PG0804+761 &   0.100      & 0.09979   &   Sy 1      & 3272 &  1 & $43.01\pm0.045$ &  1 & $8.336^{+0.087}_{-0.109}$ & $$ & &  FABS \\
     RBS1897    &   0.100       &  0.10162  &   Sy 1.5  & 2350 &  6 & $42.46\pm0.044$ &  6 & $7.702^{+0.113}_{-0.152}$ & $$ & &  FABS \\
  1H0419-577 &    0.104       & 0.10429   &   Sy 1.5   & 2580 &  6 & $43.00\pm0.044$ &  6 & $8.123^{+0.072}_{-0.087}$ & $$ & & FABSL \\
  Q1230+0115 &   0.117       &  0.11667   &   Sy 1      & 2170 &  4 & $42.93\pm0.044$ &  4 & $7.932^{+0.087}_{-0.108}$ & $$ & &   ABS \\
      Mrk876       &  0.129        &  0.12908   &   Sy 1      & 8660 &  1 & $43.02\pm0.045$ &  1 & $9.187^{+0.087}_{-0.109}$ & $8.446^{+0.165}_{-0.270}$ & 5 & FABSL \\
    VIIZw244     &   0.131       & 0.13118    & Sy 1        & 2899 &  1 & $42.37\pm0.046$ &  1 & $7.825^{+0.087}_{-0.109}$ & $$ & & FABSL \\
  PG1626+554 &  0.133       & 0.13185    & Sy 1        & 4618 &  1 & $42.74\pm0.044$ &  1 & $8.462^{+0.087}_{-0.109}$ & $$ & &    FA \\
  PG0026+129 &  0.142       & 0.14516    &  Sy 1        & 2243 &  1 & $42.85\pm0.045$ &  1 & $7.907^{+0.087}_{-0.109}$ & $8.594^{+0.095}_{-0.122}$ & 5 &   FAB \\
  PG1115+407 &  0.154567& 0.15392    &  Sy 1.0    & 2136 &  1 & $42.32\pm0.046$ &  1 & $7.533^{+0.087}_{-0.109}$ & $$ & & FABSL \\
  PG1307+085 &   0.155      & 0.15440    &  Sy 1.2   & 3860 &  6 & $42.78\pm0.044$ &  6 & $8.335^{+0.060}_{-0.069}$ & $$ & &  FABS \\
       3C273        & 0.158339 & 0.15666    &  Sy 1$^*$      & 3625 &  1 & $44.38\pm0.046$ &  1 & $9.290^{+0.087}_{-0.109}$ & $8.947^{+0.083}_{-0.103}$ & 5 &   FAB \\
  PG1202+281 &  0.1653     &  0.16533  &  Sy 1.2   & 3870 &  6 & $42.44\pm0.044$ &  6 & $8.123^{+0.080}_{-0.099}$ & $$ & &   ABS \\
  PG1048+342 &  0.167132 & 0.16681  &  Sy 1.0   & 3880 &  1 & $42.40\pm0.046$ &  1 & $8.097^{+0.087}_{-0.109}$ & $$ & &  FASL \\
  PG1116+215 &  0.1765     & 0.17470   & Sy 1.0    & 3024 &  1 & $43.53\pm0.044$ &  1 & $8.595^{+0.087}_{-0.109}$ & $$ & &  ABSL \\
  PG1121+422 &  0.225025 & 0.22409  & Sy 1.0     & 2656 &  1 & $43.18\pm0.045$ &  1 & $8.264^{+0.087}_{-0.109}$ & $$ & &    FA \\
  PG0953+415 &   0.2341    & 0.23298   &  Sy 1       & 3225 &  1 & $43.80\pm0.045$ &  1 & $8.820^{+0.087}_{-0.109}$ & $8.441^{+0.084}_{-0.104}$ & 5 &  FASL \\
 PKS1302-102 &   0.2784   & 0.27762    & Sy 1$^*$       & 3732 &  1 & $42.86\pm0.046$ &  1 & $8.355^{+0.087}_{-0.109}$ & $$ & & FABSL \\
      Ton580       &  0.290237& 0.28982    &   Sy 1$^*$       & 3400 &  6 & $42.65\pm0.044$ &  6 & $8.142^{+0.061}_{-0.071}$ & $$ & &   FAS \\
\enddata

\tablenotetext{a} {AGN properties:  redshifts ($z_{\rm QSO}$ from NED and $z_{\rm UV}$ from UV emission lines);  AGN ionization type from NED, including 
four flat-spectrum radio-loud quasars (denoted with $^*$);  line widths and inferred masses from H$\beta$  and reverberation mapping. Columns 6, 8, and 11 indicate the references from which the measurements in Columns 5, 7, and 10, respectively, were taken. }
\tablenotetext{b}{References: (1) \citet{mar03}; (2) \citet{gru99}; (3) \citet{oht07}; (4) \citet{lan08}; (5) \citet{pet04}; (6) \citet{gru04}; 
(7) \citet{den10}; (8) \citet{gri12}; (9) \citet{pet05}; (10) \citet{goo89}; (11) \citet{sti90}.}
\tablenotetext{c}{The following flags indicate which UV spectrograph was used for each object: F indicates \textit{FUSE}, A indicates COS/G130M, 
B indicates COS/G160M, S indicates \textit{IUE} short wavelength, and L indicates \textit{IUE} long wavelength.}
\tablenotetext{d}{Because no H$\beta$ luminosity was available for Mrk\,1044, we instead report the optical monochromatic continuum luminosity,
$\log\left [\lambda L_{\lambda}(\unit{5100}{\angstrom})/{\rm erg~s^{-1}}\right ]$, and mass computed using Equation 5 from \citet{ves06}.}
\end{deluxetable}

%% file: agnuvlumtable.tex
\begin{deluxetable}{lrrrrrrrrr}

\rotate
\tabletypesize{\tiny}
\tablecaption{AGN UV Luminosities\label{tab:agnuvlum}}
\tablecolumns{10}
\tablewidth{0pt}
\tablehead{
    \colhead{Object} &  \colhead{$\log\left [\frac{\lambda L_{\lambda}({\rm 1050~\AA})}{\rm erg~s^{-1}}\right ]$} &
    \colhead{$\log\left [\frac{\lambda L_{\lambda}({\rm 1350~\AA})}{\rm erg~s^{-1}}\right ]$} &
    \colhead{$\log\left [\frac{\lambda L_{\lambda}({\rm 1450~\AA})}{\rm erg~s^{-1}}\right ]$} &
    \colhead{$\log\left [\frac{L({\rm CIV})}{\rm erg~s^{-1}}\right ]$} & \colhead{$\log\left [\frac{L({\rm OVI})}{\rm erg~s^{-1}}\right ]$} &
    \colhead{$\log\left [\frac{L({\rm CII})}{\rm erg~s^{-1}}\right ]$} & \colhead{$\log\left [\frac{L({\rm MgII})}{\rm erg~s^{-1}}\right ]$} &
    \colhead{$\log\left [\frac{L({\rm HeII})}{\rm erg~s^{-1}}\right ]$} & \colhead{$\log\left [\frac{L({\rm OIII})}{\rm erg~s^{-1}}\right ]$} \
}
\startdata
     NGC4395 & $39.776\pm 0.021$ & $43.924\pm 0.001$ & $43.924\pm 0.001$ & $ $ & $36.945^{+0.029}_{-0.031}$ & $ $ & $ $ & $38.264^{+0.010}_{-0.010}$ & $38.031^{+0.013}_{-0.014}$ \\
     NGC4051 & $41.152\pm 0.013$ & $45.156\pm 0.002$ & $45.140\pm 0.002$ & $40.169^{+0.003}_{-0.003}$ & $ $ & $38.615^{+0.024}_{-0.025}$ & $39.542^{+0.025}_{-0.026}$ & $39.371^{+0.010}_{-0.011}$ & $39.009^{+0.013}_{-0.014}$ \\
     NGC3516 & $42.630\pm 0.004$ & $45.780\pm 0.002$ & $45.762\pm 0.002$ & $41.704^{+0.003}_{-0.003}$ & $ $ & $39.640^{+0.024}_{-0.025}$ & $40.789^{+0.028}_{-0.030}$ & $ $ & $ $ \\
     NGC3783 & $43.388\pm 0.001$ & $43.384\pm 0.001$ & $43.388\pm 0.001$ & $42.421^{+0.003}_{-0.003}$ & $ $ & $40.444^{+0.022}_{-0.023}$ & $41.526^{+0.010}_{-0.010}$ & $41.329^{+0.013}_{-0.013}$ & $41.188^{+0.021}_{-0.022}$ \\
     Mrk1044 & $43.424\pm 0.002$ & $46.392\pm 0.001$ & $46.391\pm 0.001$ & $42.080^{+0.002}_{-0.002}$ & $ $ & $40.920^{+0.030}_{-0.032}$ & $41.570^{+0.029}_{-0.031}$ & $41.270^{+0.018}_{-0.019}$ & $41.017^{+0.045}_{-0.050}$ \\
     NGC7469 & $43.915\pm 0.004$ & $44.336\pm 0.003$ & $44.324\pm 0.003$ & $ $ & $ $ & $41.268^{+0.044}_{-0.049}$ & $42.299^{+0.014}_{-0.014}$ & $41.592^{+0.034}_{-0.037}$ & $41.427^{+0.036}_{-0.039}$ \\
     NGC5548 & $43.798\pm 0.002$ & $43.051\pm 0.010$ & $43.052\pm 0.012$ & $42.542^{+0.007}_{-0.007}$ & $42.320^{+0.002}_{-0.002}$ & $ $ & $42.282^{+0.010}_{-0.010}$ & $40.704^{+0.028}_{-0.030}$ & $ $ \\
      Akn564 & $43.258\pm 0.002$ & $44.578\pm 0.003$ & $44.580\pm 0.003$ & $41.651^{+0.008}_{-0.008}$ & $41.449^{+0.033}_{-0.035}$ & $ $ & $41.514^{+0.045}_{-0.051}$ & $41.410^{+0.016}_{-0.017}$ & $41.028^{+0.022}_{-0.023}$ \\
      Mrk335 & $43.924\pm 0.001$ & $43.805\pm 0.002$ & $43.798\pm 0.002$ & $42.731^{+0.001}_{-0.001}$ & $42.444^{+0.012}_{-0.012}$ & $40.759^{+0.092}_{-0.116}$ & $41.990^{+0.011}_{-0.011}$ & $41.930^{+0.014}_{-0.015}$ & $41.678^{+0.024}_{-0.025}$ \\
      Mrk595 & $43.017\pm 0.013$ & $44.324\pm 0.001$ & $44.318\pm 0.001$ & $ $ & $ $ & $40.349^{+0.077}_{-0.094}$ & $ $ & $ $ & $ $ \\
      Mrk279 & $43.052\pm 0.012$ & $45.327\pm 0.045$ & $45.319\pm 0.046$ & $42.419^{+0.003}_{-0.003}$ & $41.209^{+0.035}_{-0.038}$ & $ $ & $41.482^{+0.044}_{-0.048}$ & $41.161^{+0.050}_{-0.057}$ & $41.032^{+0.080}_{-0.098}$ \\
      Mrk290 & $43.590\pm 0.002$ & $44.673\pm 0.001$ & $44.655\pm 0.001$ & $42.603^{+0.008}_{-0.008}$ & $42.355^{+0.049}_{-0.055}$ & $40.290^{+0.073}_{-0.087}$ & $41.800^{+0.021}_{-0.022}$ & $41.311^{+0.021}_{-0.022}$ & $41.194^{+0.021}_{-0.022}$ \\
      Mrk817 & $44.318\pm 0.001$ & $44.436\pm 0.003$ & $44.436\pm 0.004$ & $43.031^{+0.002}_{-0.002}$ & $42.709^{+0.005}_{-0.005}$ & $41.231^{+0.118}_{-0.163}$ & $42.549^{+0.097}_{-0.125}$ & $41.903^{+0.025}_{-0.026}$ & $41.779^{+0.028}_{-0.030}$ \\
      Mrk509 & $44.655\pm 0.001$ & $43.909\pm 0.003$ & $43.915\pm 0.004$ & $43.521^{+0.001}_{-0.001}$ & $43.506^{+0.020}_{-0.021}$ & $41.585^{+0.010}_{-0.010}$ & $43.055^{+0.016}_{-0.017}$ & $42.294^{+0.010}_{-0.011}$ & $42.302^{+0.008}_{-0.009}$ \\
  Fairall303 & $43.384\pm 0.014$ & $43.594\pm 0.002$ & $43.590\pm 0.002$ & $42.487^{+0.041}_{-0.045}$ & $41.854^{+0.028}_{-0.030}$ & $40.899^{+0.082}_{-0.101}$ & $ $ & $ $ & $ $ \\
  PG1011-040 & $44.285\pm 0.003$ & $42.600\pm 0.003$ & $42.630\pm 0.004$ & $42.513^{+0.008}_{-0.008}$ & $42.187^{+0.033}_{-0.035}$ & $41.521^{+0.059}_{-0.068}$ & $ $ & $41.703^{+0.039}_{-0.043}$ & $41.342^{+0.061}_{-0.070}$ \\
     Mrk1513 & $44.436\pm 0.004$ & $41.123\pm 0.010$ & $41.152\pm 0.013$ & $43.316^{+0.002}_{-0.002}$ & $43.124^{+0.013}_{-0.013}$ & $41.864^{+0.026}_{-0.027}$ & $42.710^{+0.049}_{-0.056}$ & $42.419^{+0.028}_{-0.030}$ & $42.299^{+0.039}_{-0.043}$ \\
  PG1229+204 & $44.324\pm 0.003$ & $43.732\pm 0.052$ & $43.757\pm 0.069$ & $42.974^{+0.014}_{-0.014}$ & $42.891^{+0.041}_{-0.045}$ & $41.364^{+0.086}_{-0.107}$ & $42.481^{+0.058}_{-0.067}$ & $41.998^{+0.037}_{-0.041}$ & $41.819^{+0.036}_{-0.039}$ \\
  MR2251-178 & $44.669\pm 0.002$ & $39.780\pm 0.019$ & $39.776\pm 0.021$ & $43.872^{+0.001}_{-0.001}$ & $43.624^{+0.015}_{-0.015}$ & $41.352^{+0.032}_{-0.034}$ & $43.110^{+0.025}_{-0.027}$ & $42.665^{+0.022}_{-0.023}$ & $42.525^{+0.011}_{-0.011}$ \\
  PG1448+273 & $43.849\pm 0.006$ & $44.407\pm 0.010$ & $44.421\pm 0.012$ & $42.425^{+0.008}_{-0.008}$ & $ $ & $41.138^{+0.089}_{-0.112}$ & $ $ & $41.815^{+0.025}_{-0.027}$ & $ $ \\
      RBS563 & $43.813\pm 0.008$ & $43.236\pm 0.002$ & $43.258\pm 0.002$ & $43.050^{+0.003}_{-0.003}$ & $42.278^{+0.018}_{-0.019}$ & $40.692^{+0.068}_{-0.081}$ & $ $ & $41.919^{+0.067}_{-0.079}$ & $ $ \\
      Mrk478 & $44.614\pm 0.013$ & $44.305\pm 0.002$ & $44.285\pm 0.003$ & $43.180^{+0.009}_{-0.009}$ & $43.181^{+0.052}_{-0.059}$ & $41.842^{+0.031}_{-0.033}$ & $42.848^{+0.014}_{-0.014}$ & $42.549^{+0.025}_{-0.027}$ & $41.989^{+0.034}_{-0.037}$ \\
     Ton1187 & $44.440\pm 0.012$ & $44.666\pm 0.002$ & $44.669\pm 0.002$ & $ $ & $42.953^{+0.018}_{-0.018}$ & $ $ & $ $ & $ $ & $ $ \\
  PG1351+640 & $44.421\pm 0.012$ & $45.458\pm 0.002$ & $45.454\pm 0.002$ & $ $ & $ $ & $41.502^{+0.051}_{-0.058}$ & $42.588^{+0.040}_{-0.044}$ & $ $ & $ $ \\
     Mrk1501 & $43.757\pm 0.069$ & $44.689\pm 0.004$ & $44.682\pm 0.005$ & $42.804^{+0.023}_{-0.024}$ & $ $ & $ $ & $42.277^{+0.040}_{-0.044}$ & $ $ & $ $ \\
  PG1411+442 & $44.580\pm 0.003$ & $45.018\pm 0.004$ & $45.000\pm 0.004$ & $ $ & $43.275^{+0.018}_{-0.019}$ & $ $ & $42.681^{+0.102}_{-0.133}$ & $42.494^{+0.068}_{-0.081}$ & $42.101^{+0.083}_{-0.102}$ \\
  PG0804+761 & $45.454\pm 0.002$ & $44.842\pm 0.003$ & $44.829\pm 0.003$ & $43.996^{+0.009}_{-0.009}$ & $43.684^{+0.035}_{-0.038}$ & $42.621^{+0.052}_{-0.060}$ & $43.679^{+0.065}_{-0.077}$ & $ $ & $ $ \\
     RBS1897 & $44.683\pm 0.002$ & $44.714\pm 0.003$ & $44.699\pm 0.004$ & $43.688^{+0.006}_{-0.006}$ & $43.420^{+0.016}_{-0.017}$ & $41.832^{+0.194}_{-0.359}$ & $42.936^{+0.153}_{-0.239}$ & $42.481^{+0.091}_{-0.116}$ & $42.143^{+0.141}_{-0.209}$ \\
  1H0419-577 & $45.012\pm 0.001$ & $45.640\pm 0.003$ & $45.648\pm 0.004$ & $44.022^{+0.002}_{-0.002}$ & $43.400^{+0.034}_{-0.037}$ & $41.978^{+0.032}_{-0.035}$ & $ $ & $42.763^{+0.118}_{-0.162}$ & $42.659^{+0.102}_{-0.133}$ \\
  Q1230+0115 & $45.235\pm 0.003$ & $45.100\pm 0.007$ & $45.092\pm 0.009$ & $43.560^{+0.004}_{-0.004}$ & $ $ & $42.714^{+0.089}_{-0.112}$ & $ $ & $ $ & $ $ \\
      Mrk876 & $45.319\pm 0.046$ & $45.690\pm 0.003$ & $45.696\pm 0.003$ & $43.983^{+0.004}_{-0.004}$ & $43.535^{+0.030}_{-0.032}$ & $41.454^{+0.091}_{-0.116}$ & $43.521^{+0.028}_{-0.030}$ & $ $ & $ $ \\
    VIIZw244 & $44.682\pm 0.005$ & $43.425\pm 0.002$ & $43.424\pm 0.002$ & $43.111^{+0.011}_{-0.011}$ & $43.036^{+0.004}_{-0.004}$ & $42.142^{+0.086}_{-0.107}$ & $ $ & $ $ & $ $ \\
  PG1626+554 & $45.000\pm 0.004$ & $43.008\pm 0.011$ & $43.017\pm 0.013$ & $43.725^{+0.006}_{-0.006}$ & $43.451^{+0.019}_{-0.020}$ & $41.900^{+0.052}_{-0.060}$ & $ $ & $ $ & $ $ \\
  PG0026+129 & $45.140\pm 0.002$ & $43.828\pm 0.007$ & $43.813\pm 0.008$ & $43.531^{+0.023}_{-0.024}$ & $42.816^{+0.030}_{-0.032}$ & $ $ & $43.051^{+0.167}_{-0.276}$ & $ $ & $ $ \\
  PG1115+407 & $44.829\pm 0.003$ & $44.705\pm 0.002$ & $44.683\pm 0.002$ & $43.168^{+0.065}_{-0.076}$ & $ $ & $42.337^{+0.058}_{-0.067}$ & $ $ & $ $ & $ $ \\
  PG1307+085 & $45.205\pm 0.002$ & $45.022\pm 0.001$ & $45.012\pm 0.001$ & $43.997^{+0.022}_{-0.024}$ & $43.570^{+0.032}_{-0.035}$ & $ $ & $ $ & $ $ & $ $ \\
       3C273 & $46.391\pm 0.001$ & $45.207\pm 0.002$ & $45.235\pm 0.003$ & $44.847^{+0.002}_{-0.002}$ & $ $ & $ $ & $44.341^{+0.005}_{-0.005}$ & $43.796^{+0.063}_{-0.074}$ & $43.681^{+0.010}_{-0.010}$ \\
  PG1202+281 & $44.170\pm 0.014$ & $45.378\pm 0.007$ & $45.361\pm 0.009$ & $43.525^{+0.006}_{-0.006}$ & $43.064^{+0.019}_{-0.020}$ & $ $ & $ $ & $ $ & $ $ \\
  PG1048+342 & $44.699\pm 0.004$ & $44.624\pm 0.010$ & $44.614\pm 0.013$ & $ $ & $ $ & $41.825^{+0.088}_{-0.111}$ & $ $ & $ $ & $ $ \\
  PG1116+215 & $45.648\pm 0.004$ & $43.823\pm 0.005$ & $43.849\pm 0.006$ & $44.278^{+0.008}_{-0.008}$ & $ $ & $42.990^{+0.056}_{-0.064}$ & $ $ & $ $ & $ $ \\
  PG1121+422 & $45.092\pm 0.009$ & $44.457\pm 0.009$ & $44.440\pm 0.012$ & $43.887^{+0.029}_{-0.031}$ & $43.709^{+0.023}_{-0.024}$ & $ $ & $ $ & $ $ & $ $ \\
  PG0953+415 & $45.762\pm 0.002$ & $45.220\pm 0.002$ & $45.205\pm 0.002$ & $44.563^{+0.011}_{-0.012}$ & $44.340^{+0.003}_{-0.003}$ & $ $ & $ $ & $ $ & $ $ \\
 PKS1302-102 & $45.696\pm 0.003$ & $44.187\pm 0.011$ & $44.170\pm 0.014$ & $ $ & $43.679^{+0.014}_{-0.014}$ & $ $ & $ $ & $ $ & $ $ \\
      Ton580 & $45.361\pm 0.009$ & $43.397\pm 0.012$ & $43.384\pm 0.014$ & $ $ & $43.878^{+0.006}_{-0.006}$ & $ $ & $ $ & $ $ & $ $ 
\enddata
\tablecomments{See Sections~2.3 and 2.4 for detailed descriptions of these measurements.}
\end{deluxetable}

%% file: agnuvwidthtable.tex
\begin{deluxetable}{lrrrrrrrrrrrr}
\rotate
\tabletypesize{\tiny}
\tablecaption{AGN UV Line Widths\label{tab:agnuvwidth}}
\tablecolumns{13}
\tablewidth{0pt}
\tablehead{
    \colhead{Object} &  \colhead{$\rm FWHM_{\rm CIV}$} &  \colhead{$\sigma_{\rm CIV}$} & \colhead{$\rm FWHM_{\rm OVI}$} &  \colhead{$\sigma_{\rm OVI}$} & \colhead{$\rm FWHM_{\rm CII}$} &  \colhead{$\sigma_{\rm CII}$} &
    \colhead{$\rm FWHM_{\rm MgII}$} &  \colhead{$\sigma_{\rm MgII}$} & \colhead{$\rm FWHM_{\rm HeII}$} &  \colhead{$\sigma_{\rm HeII}$} & \colhead{$\rm FWHM_{\rm OIII}$} &  \colhead{$\sigma_{\rm OIII}$} \\
    \colhead{} &  \colhead{(${\rm km~s^{-1}}$)}  &  \colhead{(${\rm km~s^{-1}}$)} &  \colhead{(${\rm km~s^{-1}}$)}  &  
    \colhead{(${\rm km~s^{-1}}$)} &  \colhead{(${\rm km~s^{-1}}$)}  &  \colhead{(${\rm km~s^{-1}}$)} 
    &  \colhead{(${\rm km~s^{-1}}$)}  &  \colhead{(${\rm km~s^{-1}}$)} &  \colhead{(${\rm km~s^{-1}}$)}  &  
    \colhead{(${\rm km~s^{-1}}$)} &  \colhead{(${\rm km~s^{-1}}$)}  &  \colhead{(${\rm km~s^{-1}}$)}  
}
\startdata
     NGC4395 & $   $ & $   $ & $1724\pm   6$ & $ 778\pm  66$ & $   $ & $   $ & $   $ & $   $ & $ 113\pm   3$ & $ 913\pm 179$ & $  60\pm 457$ & $1163\pm 201$ \\
     NGC4051 & $1242\pm  11$ & $2059\pm 241$ & $   $ & $   $ & $ 847\pm  38$ & $1129\pm 321$ & $1786\pm 358$ & $1573\pm 355$ & $ 398\pm  23$ & $1179\pm 184$ & $ 802\pm 413$ & $1188\pm 184$ \\
     NGC3516 & $2719\pm  27$ & $2636\pm 276$ & $   $ & $   $ & $2585\pm 134$ & $1097\pm 244$ & $4150\pm 356$ & $2704\pm 571$ & $   $ & $   $ & $   $ & $   $ \\
     NGC3783 & $2481\pm  65$ & $2604\pm 196$ & $   $ & $   $ & $3449\pm 150$ & $1463\pm 297$ & $2356\pm  55$ & $1624\pm 210$ & $ 912\pm  67$ & $1704\pm 287$ & $3485\pm 394$ & $1721\pm 288$ \\
     Mrk1044 & $1761\pm  11$ & $1441\pm 128$ & $   $ & $   $ & $1138\pm 313$ & $1388\pm 390$ & $1705\pm 416$ & $1577\pm 381$ & $1677\pm 157$ & $1281\pm 243$ & $2355\pm 927$ & $1338\pm 376$ \\
     NGC7469 & $   $ & $   $ & $   $ & $   $ & $3145\pm 890$ & $1418\pm 500$ & $3977\pm  96$ & $2902\pm 580$ & $1545\pm 345$ & $1762\pm 449$ & $3757\pm 642$ & $1942\pm 453$ \\
     NGC5548 & $1555\pm 163$ & $4318\pm 529$ & $2101\pm   3$ & $1747\pm 198$ & $   $ & $   $ & $4385\pm  87$ & $2344\pm 284$ & $ 788\pm  43$ & $ 335\pm  76$ & $   $ & $   $ \\
      Akn564 & $1263\pm  27$ & $1146\pm 213$ & $2467\pm 198$ & $ 992\pm 209$ & $   $ & $   $ & $1460\pm 354$ & $2025\pm 720$ & $ 696\pm  31$ & $ 898\pm 192$ & $1107\pm  68$ & $ 997\pm 190$ \\
      Mrk335 & $1727\pm  10$ & $1667\pm 123$ & $3420\pm 103$ & $1614\pm 287$ & $ 969\pm 443$ & $ 551\pm 295$ & $1740\pm  52$ & $1774\pm 297$ & $1431\pm 169$ & $1712\pm 307$ & $1966\pm 133$ & $1733\pm 455$ \\
      Mrk595 & $   $ & $   $ & $   $ & $   $ & $2452\pm 645$ & $1072\pm 463$ & $   $ & $   $ & $   $ & $   $ & $   $ & $   $ \\
      Mrk279 & $4266\pm  67$ & $2445\pm 182$ & $4285\pm 456$ & $2017\pm 564$ & $   $ & $   $ & $4385\pm 501$ & $3442\pm1171$ & $1541\pm 294$ & $2238\pm 692$ & $2676\pm 745$ & $2152\pm 664$ \\
      Mrk290 & $1982\pm  47$ & $3194\pm 924$ & $3776\pm1021$ & $2541\pm 664$ & $ 506\pm 385$ & $ 579\pm 227$ & $2829\pm 186$ & $3166\pm 653$ & $ 448\pm  18$ & $1701\pm 384$ & $2071\pm 504$ & $1924\pm 377$ \\
      Mrk817 & $4451\pm  56$ & $2998\pm 204$ & $5481\pm  85$ & $2312\pm 250$ & $2962\pm 670$ & $1337\pm 557$ & $3792\pm 865$ & $2636\pm1279$ & $2514\pm 491$ & $1997\pm 398$ & $4917\pm 401$ & $2118\pm 414$ \\
      Mrk509 & $3308\pm  27$ & $2778\pm 139$ & $4903\pm 250$ & $2549\pm 446$ & $2518\pm  47$ & $1069\pm 142$ & $3970\pm  94$ & $2494\pm 547$ & $2263\pm  76$ & $1673\pm 222$ & $2932\pm  75$ & $1726\pm 205$ \\
  Fairall303 & $3121\pm 563$ & $2802\pm 965$ & $2545\pm 108$ & $1356\pm 341$ & $1605\pm 319$ & $1168\pm 516$ & $   $ & $   $ & $   $ & $   $ & $   $ & $   $ \\
  PG1011-040 & $1985\pm 186$ & $1498\pm 200$ & $3339\pm 339$ & $1631\pm 470$ & $2633\pm 579$ & $1593\pm 488$ & $   $ & $   $ & $2531\pm1219$ & $1465\pm 419$ & $1814\pm 697$ & $1329\pm 418$ \\
     Mrk1513 & $2229\pm  19$ & $2063\pm 174$ & $3582\pm 203$ & $1986\pm 391$ & $1578\pm 129$ & $1182\pm 269$ & $2671\pm 326$ & $1980\pm 657$ & $2130\pm 191$ & $1948\pm 480$ & $2557\pm 267$ & $1878\pm 464$ \\
  PG1229+204 & $3648\pm 257$ & $2444\pm 467$ & $4130\pm 556$ & $2930\pm 873$ & $2145\pm 706$ & $1061\pm 400$ & $3396\pm1344$ & $2487\pm 838$ & $3859\pm 235$ & $1638\pm 394$ & $3942\pm 271$ & $1672\pm 422$ \\
  MR2251-178 & $2845\pm  17$ & $2906\pm 209$ & $2595\pm 120$ & $1553\pm 316$ & $1603\pm  92$ & $ 680\pm 158$ & $2685\pm 474$ & $3020\pm 664$ & $2375\pm 173$ & $1761\pm 385$ & $2558\pm  70$ & $1080\pm 165$ \\
  PG1448+273 & $3013\pm 113$ & $2959\pm 471$ & $   $ & $   $ & $1555\pm 656$ & $1152\pm 471$ & $   $ & $   $ & $2406\pm 374$ & $1376\pm 286$ & $   $ & $   $ \\
      RBS563 & $ 979\pm   5$ & $2383\pm 251$ & $2171\pm  39$ & $1763\pm 444$ & $ 456\pm 120$ & $ 475\pm 199$ & $   $ & $   $ & $ 280\pm  26$ & $1151\pm 480$ & $   $ & $   $ \\
      Mrk478 & $2646\pm 121$ & $2425\pm 400$ & $3614\pm 852$ & $1899\pm 666$ & $1318\pm  73$ & $ 559\pm 128$ & $2081\pm  94$ & $1594\pm 287$ & $4681\pm 282$ & $2030\pm 396$ & $2566\pm 195$ & $1084\pm 284$ \\
     Ton1187 & $   $ & $   $ & $3149\pm 297$ & $1860\pm 399$ & $   $ & $   $ & $   $ & $   $ & $   $ & $   $ & $   $ & $   $ \\
  PG1351+640 & $   $ & $   $ & $   $ & $   $ & $ 744\pm  67$ & $ 316\pm  91$ & $2873\pm 504$ & $3509\pm 995$ & $   $ & $   $ & $   $ & $   $ \\
     Mrk1501 & $3985\pm 367$ & $2453\pm 548$ & $   $ & $   $ & $   $ & $   $ & $2989\pm 454$ & $1821\pm 446$ & $   $ & $   $ & $   $ & $   $ \\
  PG1411+442 & $   $ & $   $ & $3185\pm 173$ & $1911\pm 416$ & $   $ & $   $ & $2653\pm1155$ & $2446\pm1149$ & $3069\pm 971$ & $1392\pm 509$ & $1540\pm 594$ & $1109\pm 451$ \\
  PG0804+761 & $3516\pm  98$ & $2689\pm 623$ & $3654\pm 510$ & $2056\pm 595$ & $2440\pm 802$ & $1195\pm 436$ & $2628\pm 909$ & $2546\pm 958$ & $   $ & $   $ & $   $ & $   $ \\
     RBS1897 & $1788\pm 199$ & $2983\pm 300$ & $3044\pm 113$ & $2099\pm 509$ & $1991\pm1361$ & $1140\pm 639$ & $2255\pm5893$ & $2852\pm1486$ & $3227\pm1306$ & $1653\pm 687$ & $1841\pm 985$ & $1294\pm 667$ \\
  1H0419-577 & $1744\pm   9$ & $2634\pm 213$ & $2408\pm 165$ & $1683\pm 557$ & $1533\pm 199$ & $ 833\pm 194$ & $   $ & $   $ & $1526\pm1473$ & $1138\pm 617$ & $1860\pm1375$ & $1130\pm 576$ \\
  Q1230+0115 & $3384\pm  77$ & $2910\pm 301$ & $   $ & $   $ & $2254\pm2017$ & $1230\pm 584$ & $   $ & $   $ & $   $ & $   $ & $   $ & $   $ \\
      Mrk876 & $5469\pm  72$ & $4046\pm 451$ & $5561\pm 399$ & $2574\pm 547$ & $1400\pm 248$ & $ 594\pm 236$ & $6852\pm 334$ & $2914\pm 621$ & $   $ & $   $ & $   $ & $   $ \\
    VIIZw244 & $3022\pm 213$ & $1687\pm 320$ & $4735\pm  53$ & $2034\pm 202$ & $2656\pm 255$ & $2068\pm 587$ & $   $ & $   $ & $   $ & $   $ & $   $ & $   $ \\
  PG1626+554 & $3646\pm 108$ & $2571\pm 320$ & $4501\pm 224$ & $2052\pm 431$ & $2356\pm 230$ & $1000\pm 301$ & $   $ & $   $ & $   $ & $   $ & $   $ & $   $ \\
  PG0026+129 & $2851\pm 402$ & $2853\pm 712$ & $2112\pm  17$ & $ 934\pm 321$ & $   $ & $   $ & $1388\pm1121$ & $ 942\pm 516$ & $   $ & $   $ & $   $ & $   $ \\
  PG1115+407 & $2484\pm1153$ & $1677\pm 826$ & $   $ & $   $ & $2677\pm 721$ & $1515\pm 440$ & $   $ & $   $ & $   $ & $   $ & $   $ & $   $ \\
  PG1307+085 & $4739\pm1104$ & $3184\pm 746$ & $3070\pm 454$ & $1902\pm 514$ & $   $ & $   $ & $   $ & $   $ & $   $ & $   $ & $   $ & $   $ \\
       3C273 & $3996\pm  28$ & $2802\pm 201$ & $   $ & $   $ & $   $ & $   $ & $4341\pm  61$ & $1612\pm 162$ & $4744\pm 448$ & $2047\pm 441$ & $3561\pm  97$ & $1510\pm 239$ \\
  PG1202+281 & $3758\pm  93$ & $3191\pm 441$ & $2518\pm  60$ & $1412\pm 315$ & $   $ & $   $ & $   $ & $   $ & $   $ & $   $ & $   $ & $   $ \\
  PG1048+342 & $   $ & $   $ & $   $ & $   $ & $1272\pm 866$ & $1029\pm 419$ & $   $ & $   $ & $   $ & $   $ & $   $ & $   $ \\
  PG1116+215 & $4484\pm 139$ & $2631\pm 435$ & $   $ & $   $ & $2622\pm 727$ & $1410\pm 434$ & $   $ & $   $ & $   $ & $   $ & $   $ & $   $ \\
  PG1121+422 & $2757\pm 183$ & $2336\pm 888$ & $4573\pm 275$ & $2104\pm 461$ & $   $ & $   $ & $   $ & $   $ & $   $ & $   $ & $   $ & $   $ \\
  PG0953+415 & $4002\pm 226$ & $2450\pm 414$ & $4181\pm  52$ & $2452\pm 222$ & $   $ & $   $ & $   $ & $   $ & $   $ & $   $ & $   $ & $   $ \\
 PKS1302-102 & $   $ & $   $ & $3237\pm 127$ & $1553\pm 281$ & $   $ & $   $ & $   $ & $   $ & $   $ & $   $ & $   $ & $   $ \\
      Ton580 & $   $ & $   $ & $2956\pm  56$ & $1648\pm 212$ & $   $ & $   $ & $   $ & $   $ & $   $ & $   $ & $   $ & $   $ 
\enddata
\tablecomments{See Section 2.4 for detailed descriptions of these measurements.}

\end{deluxetable}

%% file: corlumtable.tex
\begin{deluxetable}{lrrrrrrrr}

\tabletypesize{\tiny}
\tablecaption{Luminosity Correlation Matrix\label{tab:lumcor}}
\tablecolumns{9}
\tablewidth{0pt}
\tablehead{
                          $$ &        $\log{L}_{\rm H\beta}$ &  $\log\lambda{L}_{\rm 1450}$ &          $\log{L}_{\rm CIV}$ &          $\log{L}_{\rm CII}$ &         $\log{L}_{\rm MgII}$ &          $\log{L}_{\rm OVI}$ &         $\log{L}_{\rm HeII}$ &           $\log{L}_{\rm OIII}$ \\ 
}
\startdata
       $\log{L}_{\rm H\beta}$ &                                 &                           0.86 &                            0.90 &                            0.77 &                            0.82 &                            0.85 &                            0.88 &                            0.89 \\
  $\log\lambda{L}_{\rm 1450}$ &                        3.1E-13 &                                  &                           0.94 &                            0.91 &                            0.99 &                            0.91 &                            0.95 &                            0.97 \\
          $\log{L}_{\rm CIV}$ &                        2.9E-12 &                         1.6E-17 &                                  &                           0.79 &                            0.98 &                            0.92 &                            0.91 &                            0.98 \\
          $\log{L}_{\rm CII}$ &                        3.7E-06 &                         7.0E-12 &                         3.0E-06 &                                  &                           0.86 &                            0.59 &                            0.79 &                            0.82 \\
         $\log{L}_{\rm MgII}$ &                        6.2E-06 &                         6.0E-19 &                         6.8E-15 &                         1.2E-05 &                                  &                           0.91 &                            0.91 &                            0.97 \\
          $\log{L}_{\rm OVI}$ &                        2.5E-09 &                         5.3E-12 &                         7.0E-11 &                           0.013 &                         8.8E-07 &                                  &                           0.88 &                            0.93 \\
         $\log{L}_{\rm HeII}$ &                        1.4E-07 &                         3.1E-12 &                         4.0E-08 &                         0.00015 &                         2.1E-07 &                         2.4E-06 &                                  &                           0.97 \\
         $\log{L}_{\rm OIII}$  &                        7.3E-07 &                         2.6E-12 &                         2.8E-11 &                         0.00015 &                         3.4E-10 &                         4.2E-07 &                         5.9E-12 &                                 \\
\enddata
\tablecomments{Correlations among luminosities. The Spearman rank correlation coefficients, $\rho$, are reported above the diagonal. The probabilities of the correlations arising by chance, $P$, are reported below the diagonal.}
\end{deluxetable}

%% file: corwidthtable.tex
\begin{deluxetable}{lrrrrrrr}
\tabletypesize{\tiny}
\tablecaption{Line-Width Correlation Matrix\label{tab:widthcor}}
\tablecolumns{7}
\tablewidth{0pt}
\tablehead{
                          $$ &       $\rm FWHM_{\rm H\beta}$ &         $\rm FWHM_{\rm CIV}$ &         $\rm FWHM_{\rm CII}$ &        $\rm FWHM_{\rm MgII}$ &         $\rm FWHM_{\rm OVI}$ &        $\rm FWHM_{\rm HeII}$ &          $\rm FWHM_{\rm OIII}$ \\ 
}
\startdata
      $\rm FWHM_{\rm H\beta}$ &                                 &                           0.50 &                            0.12 &                            0.79 &                            0.39 &                            0.18 &                            0.68 \\
         $\rm FWHM_{\rm CIV}$ &                         0.0021 &                                  &                           0.51 &                            0.61 &                            0.60 &                            0.71 &                            0.89 \\
         $\rm FWHM_{\rm CII}$ &                           0.54 &                          0.0086 &                                  &                           0.43 &                            0.44 &                            0.45 &                            0.58 \\
        $\rm FWHM_{\rm MgII}$ &                        4.9E-06 &                          0.0033 &                           0.088 &                                  &                           0.56 &                            0.18 &                            0.72 \\
         $\rm FWHM_{\rm OVI}$ &                          0.031 &                          0.0017 &                           0.078 &                           0.023 &                                  &                           0.49 &                            0.84 \\
        $\rm FWHM_{\rm HeII}$ &                           0.41 &                         0.00051 &                           0.068 &                            0.48 &                           0.048 &                                  &                           0.45 \\
        $\rm FWHM_{\rm OIII}$  &                        0.00098 &                         1.5E-06 &                           0.024 &                          0.0012 &                         8.0E-05 &                           0.048 &                                 \\
\enddata
\tablecomments{Correlations among emission-line widths. The Spearman rank correlation coefficients, $\rho$, are reported above the diagonal. The probabilities of the correlations arising by chance, $P$, are reported below the diagonal.}
\end{deluxetable}

%% file: fitparamtable.tex
\begin{deluxetable}{lcccccc}
\tabletypesize{\footnotesize}
\tablecaption{Mass-Scaling Relationships\label{tab:fitparam}}
\tablecolumns{7}
\tablewidth{0pt}
\tablehead{
    \colhead{Variables} & \colhead{$\alpha$} & \colhead{$\beta$} & \colhead{$\gamma$}  & \colhead{$\sigma$} & \colhead{$\rho$} & \colhead{$P$} \\
   &&&&\colhead{(dex)}&& \\
    \colhead{(1)} & \colhead{(2)} & \colhead{(3)} & \colhead{(4)}  & \colhead{(5)} & \colhead{(6)} & \colhead{(7)} 
}
\startdata
	$L_{\rm CIV}$, $\rm FWHM_{CIV}$ & $ 1.47\pm 0.02$\tablenotemark{a} & [$0.53$] & [$2.00$] & 0.38 & 0.79 & 2.E-08 \\
	$L_{\rm CIV}$, $\rm FWHM_{CIV}$ & $ 1.38\pm 0.10$ & $0.45\pm 0.08$ & [$2.00$] & 0.37 & 0.86 & 2.E-11 \\
	$L_{\rm OVI}$, $\rm FWHM_{OVI}$ & $ 1.44\pm 0.02$\tablenotemark{a} & [$0.53$] & [$2.00$] & 0.43 & 0.53 &  0.002 \\
	$L_{\rm OVI}$, $\rm FWHM_{OVI}$ & $ 1.48\pm 0.12$ & $0.59\pm 0.07$ & [$2.00$] & 0.43 & 0.75 & 2.E-06 \\
	$L_{\rm MgII}$, $\rm FWHM_{MgII}$ & $ 1.84\pm 0.02$\tablenotemark{a} & [$0.53$] & [$2.00$] & 0.36 & 0.78 & 8.E-06 \\
	$L_{\rm MgII}$, $\rm FWHM_{MgII}$ & $ 1.68\pm 0.17$ & $0.46\pm 0.09$ & [$2.00$] & 0.32 & 0.79 & 4.E-06 \\
	$L_{\rm OIII}$, $\rm FWHM_{OIII}$ & $ 2.17\pm 0.02$\tablenotemark{a} & [$0.53$] & [$2.00$] & 0.47 & 0.75 & 0.0001 \\
	$L_{\rm OIII}$, $\rm FWHM_{OIII}$ & $ 1.96\pm 0.32$ & $0.45\pm 0.13$ & [$2.00$] & 0.45 & 0.78 & 5.E-05 \\
	$L_{\rm CIV}$, $\rm FWQM_{CIV}$ & $ 0.87\pm 0.02$\tablenotemark{a} & [$0.53$] & [$2.00$] & 0.36 & 0.73 & 8.E-07 \\
	$L_{\rm CIV}$, $\rm FWQM_{CIV}$ & $ 0.80\pm 0.09$ & $0.47\pm 0.07$ & [$2.00$] & 0.34 & 0.85 & 1.E-10 \\
	$L_{\rm OVI}$, $\rm FWQM_{OVI}$ & $ 1.05\pm 0.02$\tablenotemark{a} & [$0.53$] & [$2.00$] & 0.44 & 0.52 &  0.003 \\
	$L_{\rm OVI}$, $\rm FWQM_{OVI}$ & $ 1.02\pm 0.12$ & $0.53\pm 0.07$ & [$2.00$] & 0.46 & 0.77 & 6.E-07 \\
	$L_{\rm MgII}$, $\rm FWQM_{MgII}$ & $ 1.30\pm 0.02$\tablenotemark{a} & [$0.53$] & [$2.00$] & 0.32 & 0.75 & 2.E-05 \\
	$L_{\rm MgII}$, $\rm FWQM_{MgII}$ & $ 1.26\pm 0.15$ & $0.53\pm 0.08$ & [$2.00$] & 0.31 & 0.85 & 2.E-07 \\
	$L_{\rm OIII}$, $\rm FWQM_{OIII}$ & $ 1.65\pm 0.02$\tablenotemark{a} & [$0.53$] & [$2.00$] & 0.40 & 0.74 & 0.0002 \\
	$L_{\rm OIII}$, $\rm FWQM_{OIII}$ & $ 1.87\pm 0.26$ & $0.62\pm 0.10$ & [$2.00$] & 0.40 & 0.79 & 3.E-05 \\
	$L_{\rm CIV}$, $\rm \sigma_{CIV}$ & $ 1.51\pm 0.02$\tablenotemark{a} & [$0.53$] & [$2.00$] & 0.37 & 0.86 & 3.E-11 \\
	$L_{\rm CIV}$, $\rm \sigma_{CIV}$ & $ 1.62\pm 0.09$ & $0.58\pm 0.07$ & [$2.00$] & 0.29 & 0.80 & 7.E-09 \\
	$L_{\rm OVI}$, $\rm \sigma_{OVI}$ & $ 1.93\pm 0.02$\tablenotemark{a} & [$0.53$] & [$2.00$] & 0.39 & 0.68 & 4.E-05 \\
	$L_{\rm OVI}$, $\rm \sigma_{OVI}$ & $ 1.95\pm 0.11$ & $0.56\pm 0.07$ & [$2.00$] & 0.37 & 0.75 & 2.E-06 \\
	$L_{\rm MgII}$, $\rm \sigma_{MgII}$ & $ 2.07\pm 0.02$\tablenotemark{a} & [$0.53$] & [$2.00$] & 0.46 & 0.88 & 2.E-08 \\
	$L_{\rm MgII}$, $\rm \sigma_{MgII}$ & $ 2.06\pm 0.21$ & $0.56\pm 0.11$ & [$2.00$] & 0.40 & 0.76 & 1.E-05 \\
	$L_{\rm OIII}$, $\rm \sigma_{OIII}$ & $ 2.46\pm 0.02$\tablenotemark{a} & [$0.53$] & [$2.00$] & 0.59 & 0.80 & 2.E-05 \\
	$L_{\rm OIII}$, $\rm \sigma_{OIII}$ & $ 3.18\pm 0.28$ & $0.82\pm 0.11$ & [$2.00$] & 0.36 & 0.82 & 9.E-06 \\
	$\lambda L_{\rm 1450}$, $\rm FWHM_{CIV}$ & $ 0.85\pm 0.02$\tablenotemark{a} & [$0.53$] & [$2.00$] & 0.47 & 0.68 & 6.E-06 \\
	$\lambda L_{\rm 1450}$, $\rm FWHM_{CIV}$ & $ 0.86\pm 0.08$ & $0.35\pm 0.08$ & [$2.00$] & 0.43 & 0.80 & 7.E-09 \\
	$\lambda L_{\rm 1050}$, $\rm FWHM_{OVI}$ & $ 0.58\pm 0.02$\tablenotemark{a} & [$0.53$] & [$2.00$] & 0.49 & 0.52 &  0.003 \\
	$\lambda L_{\rm 1050}$, $\rm FWHM_{OVI}$ & $ 0.48\pm 0.10$ & $0.67\pm 0.09$ & [$2.00$] & 0.49 & 0.70 & 2.E-05 \\
	$\lambda L_{\rm 1450}$, $\rm FWHM_{MgII}$ & $ 0.91\pm 0.02$\tablenotemark{a} & [$0.53$] & [$2.00$] & 0.37 & 0.75 & 2.E-05 \\
	$\lambda L_{\rm 1450}$, $\rm FWHM_{MgII}$ & $ 0.87\pm 0.08$ & $0.44\pm 0.08$ & [$2.00$] & 0.32 & 0.76 & 2.E-05 \\
	$\lambda L_{\rm 1450}$, $\rm FWHM_{OIII}$ & $ 0.91\pm 0.02$\tablenotemark{a} & [$0.53$] & [$2.00$] & 0.55 & 0.65 &  0.002 \\
	$\lambda L_{\rm 1450}$, $\rm FWHM_{OIII}$ & $ 0.89\pm 0.13$ & $0.35\pm 0.12$ & [$2.00$] & 0.51 & 0.69 & 0.0008 \\
\enddata
\tablecomments{Columns~2$-$4 give the results of fits to Equation~2 using the variables given in Column~1. Column~5 is the scatter of the data around the fit. Columns~6 gives the correlation coefficient of the masses predicted by the fits with the reference masses, while Column~7 give the probability of such a correlation arising randomly. Quantities in brackets were held constant in the fit. See Section~4 for details.}
\tablenotetext{a}{Fit to Equation (2) performed with \texttt{MPFIT}.}
\end{deluxetable}